\documentclass{emulateapj}
\usepackage{epsfig}
\usepackage{amsmath}
\usepackage{natbib}
\usepackage{apjfonts}
\usepackage{graphics}
\newcommand{\beq}{\begin{equation}}
\newcommand{\eeq}{\end{equation}}
\newcommand{\bea}{\begin{eqnarray}}
\newcommand{\eea}{\end{eqnarray}}
\newcommand{\trm}[1]{\textrm{#1}}

\begin{document}

\title{Exposing the Nuclear Burning Ashes of Radius Expansion Type I X-ray Bursts}
\author{ Nevin N.~Weinberg$^{1, 2}$, Lars Bildsten$^2$, and Hendrik Schatz$^3$}
\affil{$^1$Theoretical Astrophysics, California Institute of Technology, 
Pasadena, CA 91125, nnw@tapir.caltech.edu; \\$^2$Kavli Institute for Theoretical Physics, Kohn Hall, University of California, Santa Barbara, CA 93106, bildsten@kitp.ucsb.edu; \\$^3$Department of Physics and Astronomy, National Superconducting Cyclotron Laboratory \& Joint Institute of Nuclear Astrophysics, Michigan State University, East Lansing, MI 48824, schatz@nscl.msu.edu}

\begin{abstract}
We solve for the evolution of the vertical extent of the convective region of a neutron star atmosphere during a Type I X-ray burst. The convective region is well-mixed with ashes of nuclear burning due to the short turbulent mixing time scale and its extent determines the rise time of the burst light curve. Using a full nuclear reaction network, we show that the maximum vertical extent of the convective region during photospheric radius expansion (RE) bursts can be sufficiently great that: (1) some ashes of burning are ejected by the radiation driven wind during the RE phase and, (2) some ashes of burning are exposed at the neutron star surface following the RE phase. Depending on the ignition conditions, ashes with mass number in the range $A \sim 30 - 60$ are mixed in with the ejected material. As the ejected material cools during the RE phase some of the ejected heavy-element ashes cease to be fully ionized.  In addition, those ashes that remain bound to the neutron star will temporarily reside in the photosphere after it has settled back down to the neutron star surface. Some of these surface ashes are of high enough proton number $Z$ that they are not fully ionized. We calculate the expected column density of ejected and surface ashes in hydrogen-like states and determine the equivalent widths of the resulting photoionization edges from both the wind and neutron star surface. We find that these can exceed 100 eV and are potentially detectable. A detection would probe the nuclear burning processes and might enable a measurement of the gravitational redshift of the neutron star. In addition, we find that in bursts with pure helium burning layers, protons from ($\alpha$, p) reactions cause a rapid onset of the $^{12}$C(p,$\gamma$)$^{13}$N($\alpha$,p)$^{16}$O reaction sequence.  The sequence bypasses the relatively slow $^{12}$C($\alpha$, $\gamma$)$^{16}$O reaction and leads to a sudden surge in energy production that is directly observable as a rapid ($\sim \trm{ms}$) increase in flux during burst rise.
\keywords{accretion, accretion disks --- nuclear reactions, 
nucleosynthesis, abundances --- stars: neutron --- X-rays: 
bursts}
\end{abstract}

\section{Introduction}
\label{sec:intro}

Type I X-ray bursts are produced by the unstable nuclear burning of freshly accreted hydrogen- and/or helium-rich material on the surface of a neutron star (NS) in a low-mass X-ray binary (for reviews, see \citealt{Lewin:95, Bildsten:98, Strohmayer:05}). The burst energies ($10^{39}-10^{40} \trm{ ergs}$), durations ($\sim 10 - 100 \trm{ s}$), and recurrence times (hours to days), depend strongly on the composition of the accreted matter and on the accretion rate, $\dot{M}$, which can range from $10^{-11}$ to $10^{-8} M_\odot \trm{ yr}^{-1}$.  The burst properties are also sensitive to the composition of the ashes of burning from preceding bursts \citep{Taam:80} as underscored by recent burst simulations that implement large nuclear reaction networks for the energy generation \citep{Schatz:01, BrownBA:02,Woosley:04}.

The sensitivity of the nuclear energy generation rate to temperature and density concentrates the burning during a burst to a thin layer at the base of the accreted atmosphere (e.g., \citealt{Fujimoto:81, Fushiki:87}). Since the thermal time scale there is $\sim 1 - 10 \trm{ s}$ while the dynamical time scale is $\sim 10^{-6} \trm{ s}$, the temperature gradient near the burning layer is nearly adiabatic resulting in a region of highly efficient convection. The short mixing time scale ensures that the ashes of burning are well-mixed throughout the convective region. 

The vertical extent of the convective region evolves during the burst, as demonstrated in time-dependent numerical simulations (e.g., \citealt{Joss:78, Taam:80, Ayasli:82, Woosley:84, Woosley:04}). In this paper, we carry out a thorough survey of the dependence of the convection region's extent on $\dot{M}$, the composition of the accreted material, and the pre-burst thermal state of the atmosphere. We also show how the evolution of the convective extent influences the observed burst rise times. 

We demonstrate that for photospheric radius expansion (RE) bursts, in which the super-Eddington luminosity drives a radiation-driven wind, the convective region extends out to sufficiently low pressures that ashes can be ejected by the wind. Depending on the burst parameters, the wind can eject as much as $\sim1 \%$ of the accreted mass \citep{Paczynski:86, Joss:87, Nobili:94}, corresponding to the ratio of nuclear energy release to gravitational binding energy $\sim E_{\rm nuc} / E_{\rm grav}\simeq 5\trm{ MeV nucleon}^{-1} / 200 \trm{ MeV nucleon}^{-1}$. \citet{Sugimoto:84} pointed out that such mass ejection could expose the underlying helium-rich material and result in an Eddington-limited peak flux for helium rather than hydrogen. We show that the ejected ashes may be directly detectable with spectroscopy during the RE phase and afterwards, when the photosphere, laced with heavy element ashes, settles to the NS surface.   

The convective region evolves during a burst in the following sequence of events. As the base temperature rises and the nuclear energy generation increases during the early stages of a burst, the entropy in the convective region increases. Initially, only a negligible amount of thermal energy is lost to radiation diffusing into the overlying radiative region and underlying crust. Since the time scale for radiative diffusion across the convective-radiative interface is longer than the burning time scale during these early stages, the convective region extends vertically outward to lower pressures \citep{Fujimoto:81}. Eventually, the burning rate becomes sufficiently slow that the nuclear energy flux is carried most efficiently by radiation rather than by convection. At that point the convective region recedes back to higher pressures. 

Regardless of whether the convective region is expanding outwards or receding downwards, its extent at a given time is set by the radial location in the atmosphere where the constant entropy of the convective region equals the radially increasing entropy of the overlying radiative region. Based on this argument, \citet{Joss:77} showed that the convective region never acquires a high enough entropy to reach the photosphere located at column depth $\sim 1 \trm{ g cm}^{-2}$. Nonetheless, the convective region can reach pressures $\la 1\%$ of that at the base of the accreted layer \citep{Joss:78, Taam:81,Ayasli:82, Hanawa:82,Hanawa:84}. In their simulations, \citet{Woosley:04} found that in a burst with a pure He burning layer, the peak flux exceeded the Eddington limit and the convective region extended beyond their numerical surface (i.e., the resolution limit of their grid located at a pressure $\approx 0.3\%$ that at the base). 

In this paper, we solve the time-dependent entropy equation that describes the evolving thermal structure of the atmosphere and growth of the convective region. We calculate the minimum pressure reached by the convective region (or, equivalently, the minimum column depth $y_{c, \rm min}$) for a range of burst parameters. We show that $y_{c, \rm min}$ is sensitive to the burst ignition conditions and that, in general, the larger the burst peak flux and the smaller the entropy of the pre-burst atmosphere, the smaller $y_{c, \rm min}$. Thus, $y_{c, \rm min}$ decreases with decreasing $\dot{M}$.  We find that in systems where the accreted material is helium-rich, such as 4U 1820$-$30 (see \citealt{Cumming:03} and references therein), or in systems accreting solar abundances at low $\dot M$ ($\la 10^{-9} M_\odot \trm{ yr}^{-1}$), $y_{c, \rm min} \ll y_{\rm wind}$ during RE bursts, where $y_{\rm wind}$ is the column depth above which mass is ejected by the radiation-driven wind. As a result, the wind ejects some of the nuclear burning ashes. Furthermore, some of the ashes that remain bound to the NS are exposed at the photosphere. 

In \S~\ref{sec:evolution} we describe our analytic prescription for calculating the evolution of the thermal structure of the radiative and convective regions during an X-ray burst. We explain why the evolution during the early, convective stages is sensitive to $\dot M$ and accreting composition. In \S~\ref{sec:nucleosynthesis} we describe the full nuclear reaction network we use to calculate the nucleosynthesis and nuclear energy generation. In \S~\ref{sec:results} we examine how $y_{c, \rm min}$ depends on the burst ignition conditions and explore those conditions most conducive to ash ejection and exposure. 

Since the rise time of the burst light curve is determined by the thermal diffusion time at the top of the convective zone, we also evaluate the rise time dependence on burst parameters such as $\dot M$ and accreting composition. We show these results in \S~\ref{sec:results}.

In \S~\ref{sec:detecting} we show the ash composition profile and discuss the observational consequences of heavy-element ash ejection and surface exposure. Depending on the ignition conditions, nuclei as heavy as $A\sim60$ are ejected by the wind.  We calculate the column density of ejected and surface ashes in hydrogen-like states and discuss the prospects for detecting the resulting photoionization edge features during high spectral resolution observations of RE bursts. Such features probe the nuclear burning and may constrain the NS equation of state. If bursts ignite in the ashes of previous bursts, as \citet{Woosley:04} suggest, even heavier elements are ejected. These may include some light p-nuclei \citep{Schatz:98, Schatz:01}, whose origins are not understood. We conclude in \S~\ref{sec:summary} with a summary of our work and mention the possibility of ash ejection during superbursts.

\section{Evolution of the Atmosphere During a Burst}
\label{sec:evolution}
In this section we consider the temporal evolution of the atmosphere's thermal structure during a burst. We start in \S~\ref{sec:thermal} with a general description of the atmosphere's structure and explain the boundary condition at the convective-radiative interface. The evolution of the location of the convective-radiative interface is described in \S~\ref{sec:temporal}.

\subsection{Thermal structure of the atmosphere}
\label{sec:thermal}
The NS atmosphere maintains hydrostatic equilibrium throughout the burst so that pressure varies with height as $dP /dr= -\rho g$, where $\rho$ is the density and $g$ is the surface gravity. We assume a NS mass $M = 1.4 M_\odot$ and radius $R = 10 \trm{ km}$, giving $g = (1+z) GM/R^2=2.43\times10^{14} \trm{ cm s}^{-2}$, where the gravitational redshift $z = (1 - 2 G M/Rc^2)^{-1/2} - 1 = 0.31$. Since the atmosphere is thin compared with the NS radius, $g$ is effectively constant throughout the accreted layer. Hydrostatic balance therefore yields $P = g y$, where the column depth $y$, defined by $dy = -\rho  dr$, is a convenient parameterization of the vertical spatial coordinate. We determine the extent of the convective region over a broad range of burst parameters and thus consider only one-dimensional models in our calculations. We do not account for the affect of a spreading burning front during burst rise nor the influence of rotation on the convective structure, though these effects may be important (see \citealt{Spitkovsky:02}). 

The entire atmosphere is radiative before helium burning at the base triggers the burst. The thermal profile is then described by the diffusion equation
\beq
\label{eq:heat}
\frac{dT}{dy} = \frac{3 \kappa F}{4 a c T^3},
\eeq
where $F$ is the outward heat flux and the opacity $\kappa$ has contributions from electron scattering and free-free absorption and is calculated using the approximation given by \citet{Schatz:99}. The pre-burst flux $F_0 = F_H + F_{\rm crust}$, where $F_H $ is the flux from stable hydrogen burning via the hot CNO cycle and $F_{\rm crust}$ is the flux from heat released by electron captures and pycnonuclear reactions deep in the crust \citep{Brown:98, Brown:00, Brown:04}. Following the burst ignition calculations of Cumming (2003, hereafter C03), $F_H = \epsilon_H \min[y_H, y_b]$ where $\epsilon_H = 5.8 \times 10^{13} \trm{ ergs g}^{-1} \trm{ s}^{-1} (Z/ 0.01)$ is the hot CNO energy production rate for a CNO mass fraction $Z$, $y_H$ is the column depth of the layer that is burning hydrogen, and $y_b$ is the column depth at the base. For typical ignition conditions $y_b \simeq 3 \times 10^8 \trm{ g cm}^{-2}$.  For a given local accretion rate $\dot{m}$ (in units of $\trm{g cm}^{-2} \trm{ s}^{-1}$) and accreted hydrogen mass fraction $X_0$, the hydrogen burning depth is $y_H = 6.8 \times 10^8 \trm{ g cm}^{-2} (\dot{m} / 0.1 \dot{m}_{\rm Edd}) (0.01 / Z) (X_0/0.71)$. Here $\dot{m}_{\rm Edd} = 2 m_p c / (1+X_0)R\sigma_{\rm Th} = 8.8 \times 10^4 \trm{ g cm}^{-2} \trm{ s}^{-1} (1.71 / [1 + X_0])$ is the local Eddington accretion rate where $m_p$ is the proton mass, $c$ the speed of light, and $\sigma_{\rm Th}$ the Thomson scattering cross section. For $\dot m$ smaller than the critical accretion rate $\dot {m}_{\rm crit} \simeq 0.04 \dot{m}_{\rm Edd}$, $y_H < y_b$ and there is enough time to burn all the hydrogen at the base before the helium burning becomes unstable. The burst then ignites in a pure helium layer. As in C03, we assume $F_{\rm crust} =  \dot{m} Q_{\rm crust} = 10^{21} \trm{ ergs cm}^{-2} \trm{ s}^{-1}  \dot{m}_4 Q_{0.1}$ where  $\dot{m}_4 = \dot m / 10^4 \trm{ g cm}^{-2} \trm{ s}^{-1}$ and $Q_{\rm crust} = 0.1 Q_{0.1} \trm{ MeV nucleon}^{-1} \simeq Q_{0.1} 10^{17} \trm{ ergs g}^{-1}$ is the energy per nucleon released in the crust from pycnonuclear and electron capture reactions that escapes from the surface. The pre-burst flux $F_{21} = F_0 / 10^{21} \trm{ ergs cm}^{-2} \trm{ s}^{-1}$ is thus,
\beq
\label{eq:F_0}
F_{21}  =  \dot{m}_4 Q_{0.1} 
+ \min\left[37 X_0 (1+X_0)\dot{m}_4 , 6 Z_{0.01} y_8 \right],
\eeq
where $Z_{0.01} = Z / 0.01$ and $y_8 = y_b / 10^8 \trm{ g cm}^{-2}$.

The thermal evolution of the NS atmosphere during a burst is described by the entropy equation
\beq
T\frac{ds}{dt} = \frac{dF}{dy} + \epsilon, 
\label{eq:entropy}
\eeq
where $\epsilon$ is the energy release rate from nuclear burning. During the burst the entropy grows with time due to  nuclear burning and we neglect the advective accretion flow. Therefore, $T ds = C_p dT$ where $C_p$ is the specific heat at constant pressure. Integrating equation (\ref{eq:entropy}) over column depth then gives
\beq
\label{eq:cpt}
\int_{y_1}^{y_2} C_p \frac{dT}{dt} \, dy =  F(y_2) - F(y_1) + \int_{y_1}^{y_2} \epsilon \, dy.
\eeq
We assume the atmosphere is composed of two regions: a completely convective region between $y_c < y < y_b$, and a completely radiative region for $y < y_c$. During the burst rise, $y_b$ is constant while $y_c$ evolves from an initial value $y_c = y_b$ to a minimal value $y_c = y_{c, {\rm min}}$ and finally back to $y_c = y_b$.   Demarcating the atmosphere in this way is reasonable given that the convective eddies are highly subsonic over most of the convective zone, i.e., near the base $v_{\rm conv} \simeq (F/\rho)^{1/3} \sim 10^7 \trm{ cm s}^{-1} \ll c_{\rm s} \simeq (g y_b / \rho)^{1/2} \sim 2\times10^8 \trm{ cm s}^{-1}$. Near the top of the convective zone, convection becomes inefficient and $v_{\rm conv} \sim c_s$.   

The thermal profile in the radiative region satisfies equation (\ref{eq:heat}).  Since the radiative region is composed primarily of freshly accreted hydrogen and/or helium, the main opacity is Thomson scattering $\kappa \simeq \kappa_{\rm es} \simeq \sigma_{\rm Th}(1+X)/2 m_p$ (there are corrections to $\kappa_{\rm es}$ due to relativistic electrons and degeneracy). The opacity varies only slightly with column depth so that over much of the radiative region $d \ln T / d \ln y \simeq 1/4$. For mixed hydrogen/helium accretion at $\dot{m} < \dot{m}_{\rm crit}$,  a pure helium layer develops in the region $y_H < y < y_b$. In this region $F_H = 0$ and since $F_{\rm crust}$ is small at low $\dot{m}$, the pre-burst profile there is nearly isothermal.

\begin{deluxetable}{llllcc}
\tablecaption{Burst ignition models}
\tablewidth{0pt}
\tablehead{
\colhead{Model} & \colhead{$y_8$} & \colhead{$X_0$} & \colhead{$Z_{\rm CNO}$} & \colhead{$Q_{\rm crust}$} & \colhead{network}
}
\startdata
\label{tab:1}
He[$\dot m$] & 3.0 & 0.0 & 0.01 & 0.1 & full\\
He0.1$\alpha$ & 3.0 & 0.0 & 0.01 & 0.1 & $\alpha$-only\\
He0.1b & 5.0 & 0.0 & 0.01 & 0.1 & full\\
He0.1Q & 3.0 & 0.0 & 0.01 & 0.2 & full\\
HHe[$\dot m$] & 3.0 & 0.71 & 0.01 & 0.1 & full\\
HHe0.01X & 3.0 & 0.1 & 0.01 & 0.1 & full\\
HHe0.01XZ & 3.0 & 0.1 & 0.0001 & 0.1 & full\\
\enddata
\tablecomments{Col. (1): Model name, where [$\dot m$] denotes the accretion rate in units of $\dot m_{\rm Edd}$. We consider models spanning the range $0.01 \dot m_{\rm Edd}$ to $0.2 \dot m_{\rm Edd}$. Col. (2): Ignition column depth $y_8 = y_b / 10^8 \trm{ g cm}^{-2}$. Col. (3): Accreted hydrogen fraction $X_0$. Col. (4): CNO mass fraction $Z_{\rm CNO}$. Col. (5): $Q_{\rm crust}$ in units of MeV nucleon$^{-1}$. Col. (6): Reaction network used.}
\end{deluxetable}

Although the entropy in the convective region grows with time, at a given instant it is nearly spatially constant. This, in addition to the subsonic motion of the convective eddies, suggests that the thermal profile in the convective region very nearly follows an adiabat so that $(d \ln T / d \ln y)_{\rm conv}= (d \ln T / d \ln y)_{\rm ad} \equiv n(y) $, i.e., $T(y_c<y<y_b) = T_b (y/y_b)^{n(y)}$, where $T_b = T_b(t)$ is the temperature at the base and the adiabatic index $n(y)$ varies with column depth. We define the column depth of the convective-radiative interface $y_c$ as the location where the density of the radiative solution just exceeds that of the convective solution (i.e, neutral buoyancy criterion). 

For the equation of state we use the interpolation formulae of \citet{Paczynski:83} to account for partially degenerate electrons. Using his notation, the specific heat and adiabatic index are given by
\bea
 C_p & = & \frac{1}{\rho T} \left[\frac{3}{2} P_i + 12 P_r  + \frac{P_{\rm end}^2}{(f-1) P_e} + \frac{P\chi_{T}^2}{\chi_{\rho}} \right], \\
n & = & \frac{P}{C_p \rho T} \frac{\chi_{T}}{\chi_{\rho}},
\eea
where $P_i$, $P_r$, and $P_e = (P_{\rm end}^2 + P_{\rm ed}^2)^{1/2} $ are the pressure due to ions, radiation, and electrons, respectively, $P_{\rm end}$ and $P_{\rm ed}$ are an approximation to the degenerate and non-degenerate components of the electron pressure, $P = P_i + P_e + P_r$, $f = d \ln P_{\rm ed} / d \ln \rho$, and
\bea
\chi_{T} & \equiv & \left(\frac{\partial \ln P}{\partial \ln T}\right)_{\rho} = \frac{1}{P}\left[P_i + 4 P_r + \frac{P_{\rm end}^2}{P_e}\right], \\
\chi_{\rho} & \equiv & \left(\frac{\partial \ln P}{\partial \ln \rho}\right)_{T}  =  \frac{1}{P}\left[P_i + \frac{P_{\rm end}^2 + f P_{\rm ed}^2}{P_e}\right].
\eea
At burst onset the pressure is nearly that of an ideal gas and $n \simeq 2/5$ while at late times radiation pressure contributes significantly, which in the limit $P=P_r$ gives $n = 1/4$.

\subsection{Temporal evolution of the thermal structure}
\label{sec:temporal}

\begin{figure}
\epsscale{1.0}
\includegraphics[bb= 10 50 580 481, angle=-90,scale=0.42]{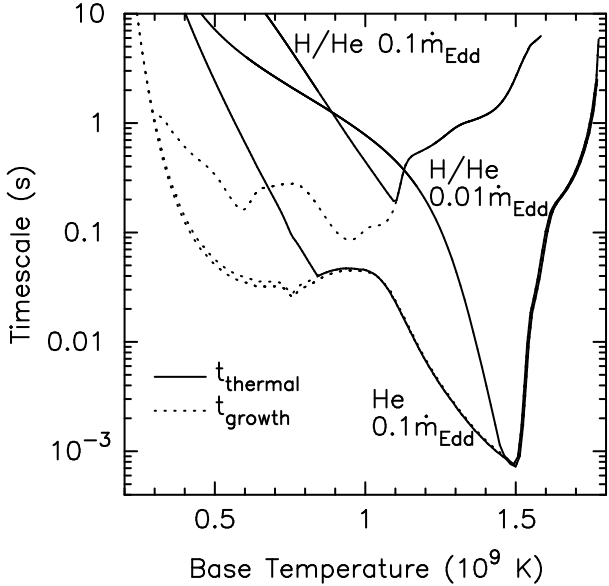}
\caption{Thermal timescale at the convective-radiative interface $t_{\rm th}(y_c)$ ({\it solid line}) and the growth timescale of the convective zone $t_{\rm gr} \equiv dt / d\ln T_b$ ({\it dotted line}) as a function of the base temperature $T_b$ for pure helium accretion ($X = 0$, $Y = 0.99$, $Z = 0.01$) at $\dot{m} / \dot{m}_{\rm Edd} =$ 0.1 and solar abundance accretion ($X = 0.71$, $Y = 0.28$, $Z = 0.01$) at $\dot{m} / \dot{m}_{\rm Edd} =$ 0.1 and 0.01. The convective zone reaches a minimum pressure approximately when the equality $t_{\rm th} = t_{\rm gr}$ is first satisfied. \label{fig:timescale}}
\end{figure}

\begin{figure*}
\includegraphics[bb=187 65 513 801, angle=-90,scale=0.8]{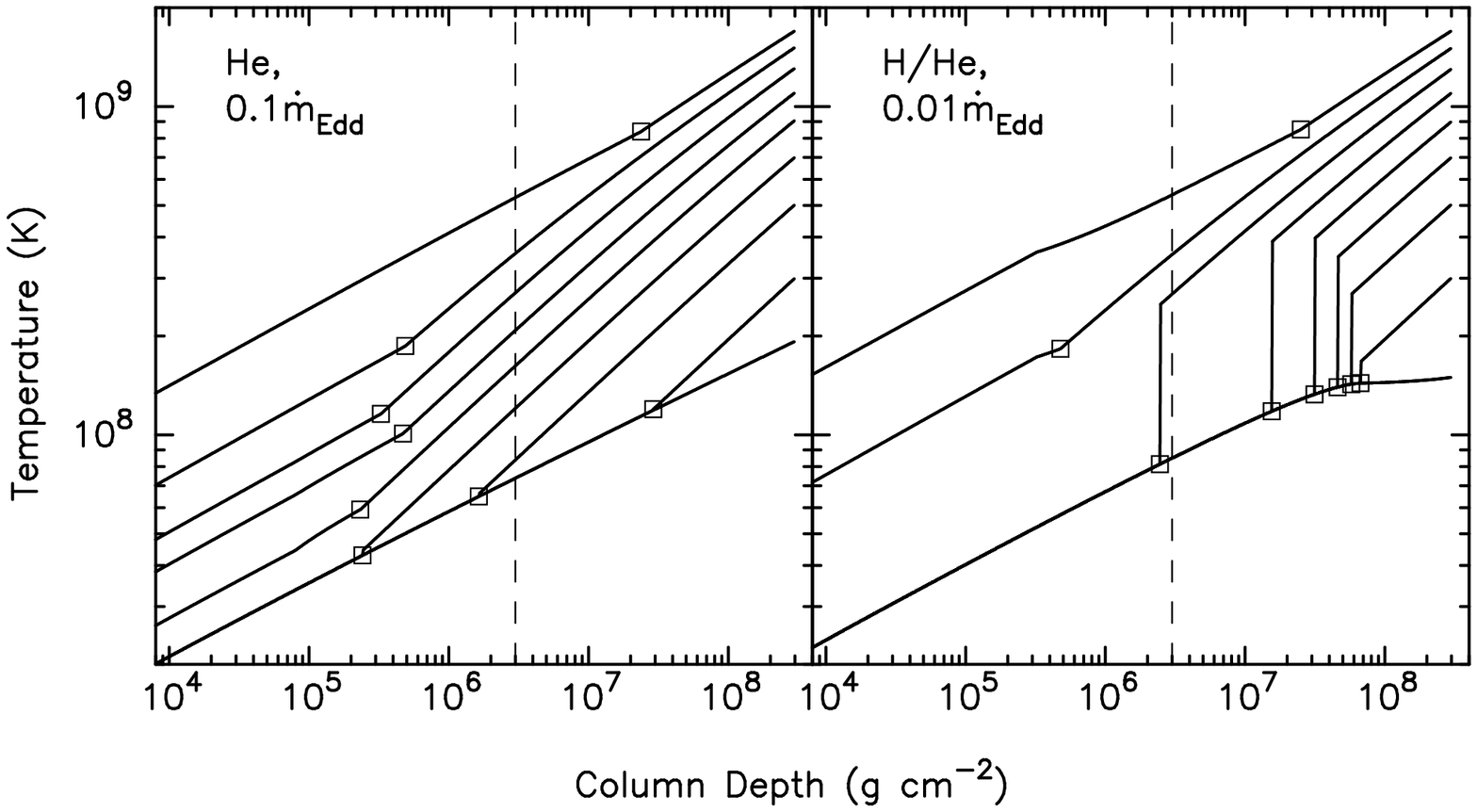}
\caption{Evolution of the temperature profile for pure helium accretion at $\dot{m} / \dot{m}_{\rm Edd} =$ 0.1  ({\it left panel}) and solar abundance accretion at $\dot{m} / \dot{m}_{\rm Edd} =$ 0.01 ({\it right panel}) at nine different times corresponding to $T_b / 10^9 \trm{ K} = T_0$, 0.3, 0.5, 0.7, 0.9, 1.1, 1.3, 1.5, and 1.7, where $T_0$ is the initial base temperature set by the radiative solution with $F_0 = F_H+F_{\rm crust}$. The open squares mark the convective-radiative interface $y_c$. The vertical {\it dashed lines} denote the column depth where $y = 0.01y_b$, corresponding approximately to the column depth $y_{\rm wind}$ below which material is ejected by the radiative wind of an RE burst.\label{fig:Ty}}
\end{figure*}

The evolution of the convective extent $y_c$ depends on the rate at which the base temperature rises, $dT_b/dt$, and the rate at which the thermal energy of the overlying radiative region increases. The rate of temperature change in the convective region is $dT/dt = (y /y_b)^n \left[dT_b/dt + T_b \ln(y/y_b)dn/dt\right]$. The second term is negligible compared with the first term so that by equation (\ref{eq:cpt}),
\beq
\label{eq:dTbdt}
\frac{d T_b}{dt} = \frac{ \mbox{\LARGE
$\int_{\mbox{\footnotesize$y_c$}}^{\mbox{\footnotesize$y_b$}}$}
  \epsilon \, dy + F_H + F_{\rm crust} - F_{\rm loss}(y_c)} {\mbox{\Large $\int_{\mbox{\footnotesize $y_c$}}^{\mbox{\footnotesize $y_b$}}$} C_p (y/y_b)^{n(y)} \, dy },
\eeq 
where $F_{\rm loss}(y_c)$ is the radiative flux escaping from the convective region into the overlying radiative region. Physically, the rate at which $T_b$ changes is determined by the competition between the net energy input into the convective region (i.e., the energy generated by nuclear burning and crustal heating minus the energy lost to radiation) and the energy expended in heating up the growing convective region. We describe our full nuclear reaction network in \S~\ref{sec:nucleosynthesis}.

We use mixing length theory to estimate $F_{\rm loss}(y_c)$, which gives the fraction of the total flux transported by convection at column depth $y$ \citep{Hansen:94},
\beq
\label{eq:mlt}
\frac{F_{\rm conv}}{F} = \frac{\nabla_{\rm rad} - \nabla_{\rm ad}}{\nabla_{\rm rad}}
\left(1 - \frac{1}{Nu}\right),
\eeq
where $\nabla_{\rm ad} \equiv n(y)$, $\nabla_{\rm rad} = 3 \kappa F y / 4 a c T^4$, and $Nu = Nu(y)$ is the Nusselt number describing the efficiency of convection. During the burst, convection is very efficient over most of the convective region and $Nu \gg 1$. Because mixing length theory only provides an order of magnitude estimate of $F_{\rm loss}(y_c)$, we introduce a scaling prefactor $\lambda$ to parameterize the uncertainty in its exact value. We thus have $F_{\rm loss}(y_c) = F - F_{\rm conv}(y_c) \approx  \lambda F \nabla_{\rm ad} / \nabla_{\rm rad} = \lambda 4ac T_c^4  \nabla_{\rm ad} / 3 \kappa y_c$ where $T_c \equiv T(y_c)$. We assume $\lambda = 1$ throughout, though as we show below, $y_{c, \rm min}$ is sensitive to the value of $\lambda$.

In Figure~\ref{fig:timescale} we show the evolution of the growth time scale of the convective zone, $t_{\rm gr} \equiv dt / d\ln T_b$, and the thermal time scale at the base of the radiative zone, $t_{\rm th} = C_p T_c y_c / F_{\rm loss}(y_c)$. During the early stages of a burst, the convective-radiative interface is located at large column depths and $t_{\rm th} \gg t_{\rm gr}$. Thus, the radiative region cannot thermally adjust to the growing convective region and the thermal profile in the radiative region is unchanged from the pre-ignition profile (see \citealt{Hanawa:82}). The initial entropy of the atmosphere, which is set by $F_0 = F_H + F_{\rm crust}$, is therefore important in determining the convective evolution.\footnote{When the density and temperature first get high enough for helium to ignite the burning time scale may be longer than or comparable to the thermal time scale. The flux from this early burning may change the pre-burst profile of the radiative region slightly, though eventually the burning becomes non-linear and the burning time becomes much shorter than the thermal time. To examine how this might affect the growth of the convective region we computed $y_c(t)$ assuming an artificially high pre-burst flux (e.g., $5 \times F_0$). We find that this effect may increase $y_{c, \rm min}$ by as much as a factor of 3.}  

Eventually the convective region reaches a low enough column depth that $t_{\rm th} = t_{\rm gr}$. The radiative flux $F_{\rm loss}(y_c)$ can then finally diffuse through the entire radiative region without being overtaken by the growing convective region. Some of this flux will heat the radiative region, while the remainder escapes through the photosphere. The subsequent evolution of the convective-radiative interface is determined by the column depth at which $t_{\rm th} = t_{\rm gr}$, i.e., the radiative region  continuously adjusts to $F_{\rm loss}(y_c)$, which varies due to changes in the burning rate.

Figure~\ref{fig:timescale} shows results for accretion of pure helium at a rate $\dot{m}/ \dot{m}_{\rm Edd} = 0.1$ and solar abundance accretion at a rate $\dot{m}/ \dot{m}_{\rm Edd} = 0.01$ and 0.1 (corresponding to models He0.1, HHe0.01, and HHe0.1 as described in Table \ref{tab:1}).  The pure He model is similar to the ignition models found by C03 in fits to the burst properties of 4U 1820-30.  For hydrogen-rich accreted material, the lower the $\dot m$ the shorter the growth time at a given $T_b$. This is because the mass fraction of helium at the burning layer, and therefore the triple alpha energy generation rate $\epsilon_{3\alpha} \propto Y^3$, increases as $\dot m$ decreases.

In models with He-rich burning layers, $t_{\rm gr}$ decreases sharply at $T_b \ga 10^9 \trm{ K}$ due to the sudden rise in energy generation $\epsilon$ when the $^{12}$C(p,$\gamma$)$^{13}$N($\alpha$,p)$^{16}$O bypass first takes over the reaction flow (see \S~\ref{sec:nucleosynthesis}). When $T_b \simeq 1.5 \times 10^9 \trm{ K}$, $t_{\rm gr}  = t_{\rm th} \sim 10^{-3} \trm{ s}$, resulting in millisecond burst rise times (\S~\ref{sec:lightcurve}).

In Figure~\ref{fig:Ty} we show the temperature profile at different stages for the pure He $\dot{m}/ \dot{m}_{\rm Edd} = 0.1$ model and the mixed H/He $\dot{m}/ \dot{m}_{\rm Edd} = 0.01$ model. For models accreting hydrogen, there is a large compositional contrast between the helium-rich matter that is burning and the outer hydrogen-rich material.\footnote{Since the particle diffusion time scale is much longer than $t_{\rm gr}$, there is not adequate time for semi-convection to develop. We therefore approximate the composition gradient as a discontinuous step function.} As we demonstrate momentarily, the contrast inhibits the outward progress of the convective zone because the burning material must get even hotter to become buoyant in the overlying hydrogen-rich envelope. {\it Thus, the minimum column depth reached by the convective zone $y_{c, \rm min}$ is significantly smaller for helium accretors as compared with hydrogen accretors, even when the latter accretes at $\dot{m} < \dot{m}_{\rm crit}$ and thus also burns in a pure helium environment} (see also \citealt{Cumming:00}). This difference in convective zone evolution between the hydrogen and helium accretors influences $t_{\rm th}(t)$ (see Figure \ref{fig:timescale}) and therefore affects the light curves during burst rise, as we show in \S~\ref{sec:lightcurve}. 

 To appreciate how the compositional contrast influences the convective evolution, first consider the very early stages of the evolution, when $t_{\rm th} \gg t_{\rm gr}$. As noted in \S~\ref{sec:thermal}, we define the convective-radiative interface as the location of neutral buoyancy, where the density of the radiative solution just equals that of the convective solution. The density is therefore continuous across the interface, even in the presence of a compositional contrast. Hydrostatic equilibrium ensures the pressure is also continuous. Thus, to accomodate the jump in mean molecular weight across the interface, the temperature must increase by a fractional amount $(T_c - T_r) / T_c \approx (\mu_c - \mu_r) / \mu_c$, where $T_c$ and $\mu_c$ are the temperature and mean molecular weight on the convective-side of the interface, $T_r$ and $\mu_r$ are those on the radiative-side, and the equality assumes gas pressure dominates. A thermal wave propagates outward from the interface, but only travels a distance $\approx h(t_{\rm gr} / t_{\rm th})^{1/2}$, where $h$ is the scale height at $y_c$, before being outrun by the growing convective zone. Thus, during the very early stages of the evolution, when $t_{\rm th} \gg t_{\rm gr}$, the time-dependent solution yields a very sharp temperature gradient. We approximate the gradient as a jump as long as $t_{\rm th} > t_{\rm gr}$, as shown in Figure ~\ref{fig:Ty}. Our approximation becomes inaccurate as the convective zone extends out to lower pressures and $t_{\rm th} \rightarrow t_{\rm gr}$, as then the thermal wave can ``get ahead'' of the growing convective zone. This approximation thus prevents us from resolving the very early light curve of the burst rise. We resolve the light curve well-before the luminosity exceeds the accretion luminosity, however. 

Once $t_{\rm th} = t_{\rm gr}$, there is enough time for heat diffusion (via radiation) to smooth out the temperature gradient over a scale height, and the radiative region heats up. During this stage the convective zone returns to higher pressure and the compositional contrast between burnt and unburnt matter is at a fixed location $y = y_{c, \rm min}$ and is thus entirely within the radiative zone. The discontinuous change in opacity at $y_{c, \rm min}$ results in a change of slope of the radiative temperature profile at that location, as can be seen in Figure ~\ref{fig:Ty}.

We now obtain a rough estimate of $y_{c, \rm min}$ by setting $t_{\rm th} = t_{\rm gr}$ and solving for $y_c$ given the pre-burst thermal profile set by $F_0$. Using the mixing length theory relation for $F_{\rm loss}(y_c)$ and assuming an adiabatic profile in the convective zone so that $T_c = T_b(y_c/y_b)^n$ yields $t_{\rm th} = 3 \kappa C_p y_b^{3n} y_c^{2-3n} / 4 a c \lambda n T_b^3$. Then, since $t_{\rm gr} \approx C_p T_b / \epsilon$, we have, upon setting $t_{\rm th} = t_{\rm gr}$, that 
\beq
y_c = \left(\frac{4 a c \lambda n T_b^4}{3 \epsilon \kappa y_b^{3n}}\right)^{\textstyle \frac{1}{2-3n}}.
\eeq
At $y_c$ the density and pressure are continuous and at $y_{c, \rm min}$ the pressure is nearly that of an ideal gas. We can thus eliminate $T_b$ from the above expression using $T_c = \mu_c T_{\rm r}(y_c) / \mu_r$ and $T_{\rm r}(y_c) \simeq (3 \kappa y_c F_0 / ac)^{1/4}$. Solving for $y_c$ we get,
\bea
y_{c, \rm min} & \simeq & \left(4 \lambda n  y_b^n \left(\frac{\mu_c}{\mu_r}\right)^4 \frac{F_0}{\epsilon}\right)^{{\textstyle \frac{1}{1+n}}} \\ & \simeq &
2 \times 10^4 \trm{ g cm}^{-2} F_{21}^{5/7} \left(\frac{\mu_c}{\mu_r}\right)^{20/7}  Y_b^{-15/7} \nonumber \\ & \times & \left(\frac{y_b}{3\times10^8 \trm{ g cm}^{-2}}\right)^{2/7} \left(\frac{\epsilon_{18} / Y_b^3}{3.0}\right)^{-5/7}, \nonumber
\eea
where the dependence on $\dot m$ is obtained by substituting equation (\ref{eq:F_0}) for $F_{21} = F_0 / 10^{21} \trm{ ergs cm}^{-2} \trm{ s}^{-1}$. In the lower expression, $Y_b$ is the helium mass-fraction in the burning layer and we assumed $n = 2/5$, $\lambda = 1$, and an energy generation rate $\epsilon_{18} = \epsilon / 10^{18} \trm{ ergs g}^{-1} \trm{ s}^{-1}$. For helium burning, $\epsilon_{18} / Y_b^3 = 5.3\times10^2 \rho_6^2 \exp(-4.4 / T_9) / T_9^3$ \citep{Hansen:94} where $T_9 = T_b / 10^9 \trm{ K}$ and $\rho_6 = \rho_b / 10^6 \trm{ g cm}^{-3}$. The estimate and its scaling with burst parameters agrees reasonably well with our full evolution calculations presented in \S~\ref{sec:results}. Note the strong dependence on compositional contrast between the accreted material and the ashes. 

As we show in \S~\ref{sec:results}, the pure helium accretion models and the solar abundance accretion models with $\dot m \la 0.05 \dot m_{\rm Edd}$ achieve a super-Eddington luminosity that drives a radiative wind capable of ejecting material located at column depths $y < y_{\rm wind} \simeq 0.01 y_b$.  Since the convective zone reaches pressures $y_{c, {\rm min}} \ll y_{\rm wind}$ in these models, one expects ashes of burning to be amongst the wind ejecta. 

\section{Nucleosynthesis from Pure Helium Burning and Mixed Hydrogen and Helium Burning}
\label{sec:nucleosynthesis}

\begin{figure*}[!ht]
\includegraphics[bb= -5 0 0 221, angle=0, scale=1.1]{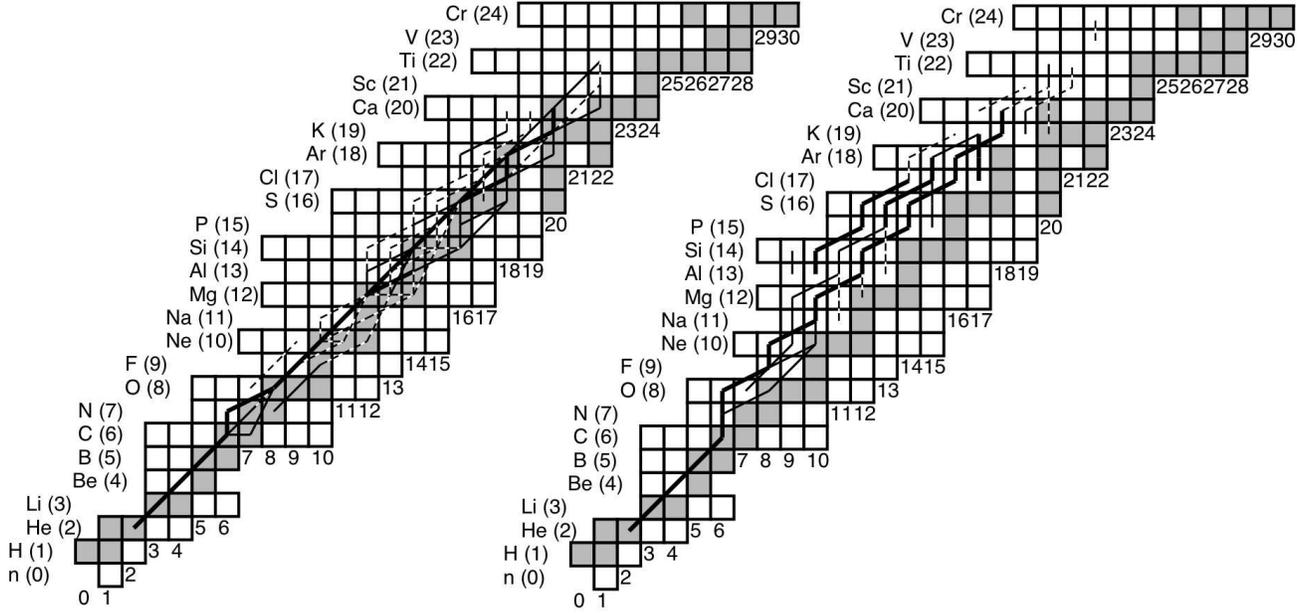}
\caption{Time-integrated net reaction flow for model He0.1 ({\it left grid}) 
and model HHe0.1 ({\it right grid}). The thickness of each line indicates the strength 
of the reaction flow relative to the 3$\alpha$ reaction: more than 10\% flow 
({\it thick solid line}), 1\%$-$10\% flow ({\it thin solid line}), and 0.1\%$-$1\% flow 
({\it dashed line}). Each square stands for a proton stable nucleus and filled squares are 
stable nuclei. Shown are all nuclei below Mn that are included in the reaction network. Note that the net flows for (p, $\gamma$) reactions can be very inaccurate when both the (p, $\gamma$) rate and the inverse ($\gamma$, p) rate are very large.
\label{fig:Flow}}
\end{figure*}

\begin{figure*}
\includegraphics[bb= 120 60 500 281, angle=-90,scale=0.8]{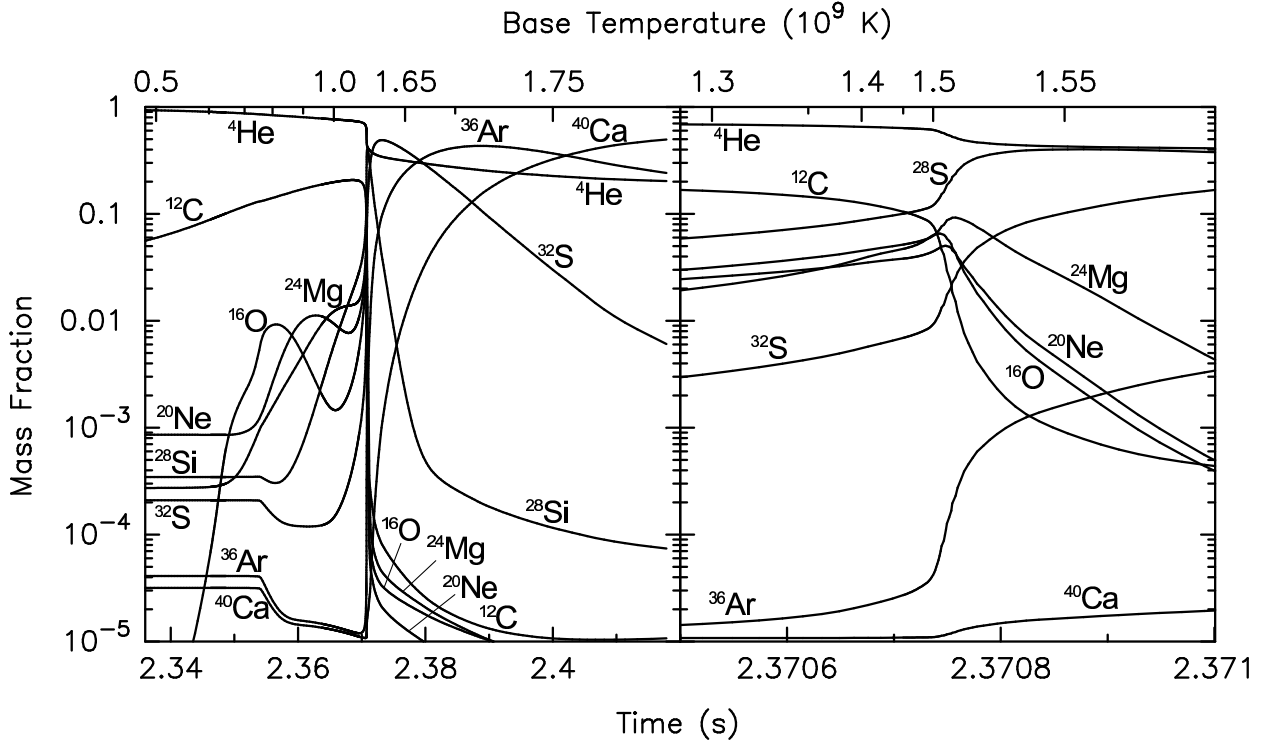}
\caption{ Mass fraction of various isotopes as a function of time
  ({\it bottom axis}) and base temperature ({\it top axis}) for model
  He0.1. The {\it left panel} shows the full evolution and the {\it
    right panel} zooms in on the time during which the
  $^{12}$C(p,$\gamma$)$^{13}$N($\alpha$,p)$^{16}$O bypass takes over
  the reaction flow from $^{12}$C to $^{16}$O. Only those isotopes
  whose peak mass fraction exceeds 0.05 are
  shown. \label{fig:AbundFlow1}}
\end{figure*}

The full nuclear reaction network used in this work contains 394
nuclei from H to Sr ranging from the proton drip line to a few mass
units beyond stability and accounts for all relevant proton and
$\alpha$ particle induced reactions and the corresponding inverse
processes. The reaction rates have recently been updated and a full
discussion will be presented in a forthcoming paper. Here we are only
concerned with the nucleosynthesis in the early, convective stages of
an X-ray burst.  The relevant nuclear reactions involve only nuclei
lighter than iron and we restrict the following discussion to those
nuclei.  When no experimental information is available, reaction rates
are taken from sd-shell model calculations \citep{Herndl:95}, fp-shell
model calculations \citep{Fisker:01}, or, the statistical model
NON-SMOKER \citep{Rauscher:00}.  Experimental rates were taken from
the NACRE \citep{Angulo:99} and \citet{Iliadis:01} compilations with
some exceptions. As far as the reaction rates relevant to this work
are concerned, these exceptions include: the 3$\alpha$-reaction where
we used the new experimental rate from \citet{Fynbo:05}, though for
the temperatures relevant here this rate is identical to the one in
\citet{Caughlan:88}. For $^{12}$C($\alpha$,$\gamma$) we use the
reaction rate by \citet{Buchmann:96}. The $^{14}$O($\alpha$,p)$^{17}$F
reaction is from \citet{Hahn:96} and the $^{13}$N(p,$\alpha$)$^{16}$O
and $^{24}$Mg($\alpha$,$\gamma$)$^{28}$Si reactions are from
\citet{Caughlan:88}.  Weak interaction rates were taken from
\citet{Fuller:82} when available, otherwise ground state $\beta$-decay
rates from Nubase \citep{Audi:03} were employed.  Weak interactions,
however, do not play a role in any of our models.

We calculate the nucleosynthesis and nuclear energy generation only at
the base of the convective zone $y_b$. To estimate the integral of
energy generation over column depth in the calculation of $dT_b / dt$
and $t_{\rm gr}$ (see equation \ref{eq:dTbdt}), we first take the
ratio of $\int \epsilon(T_b, y) dy$ to $\epsilon(T_b, y_b) y_b$
assuming only $\alpha$-capture reactions. We find that to a good
approximation the ratio is constant with temperature and we use this
constant factor to rescale $\epsilon(T_b, y_b) y_b$ calculated using
the full network. Our initial composition is the composition at $y_b$
at ignition and we neglect compositional changes due to convective
mixing as the convection zone moves outward. Such mixing might be
relevant for the H/He accretion models where the convection zone at
later stages extends into hydrogen rich regions.  However, the
effect is probably comparable to an increase in accretion rate, which
also results in a larger hydrogen abundance.  Therefore, our neglect
of convective mixing might simply translate to a slight underestimate
of the accretion rate in our sequence of calculated models.

\begin{figure}
\includegraphics[bb= 0 20 570 061, angle=-90,scale=0.46]{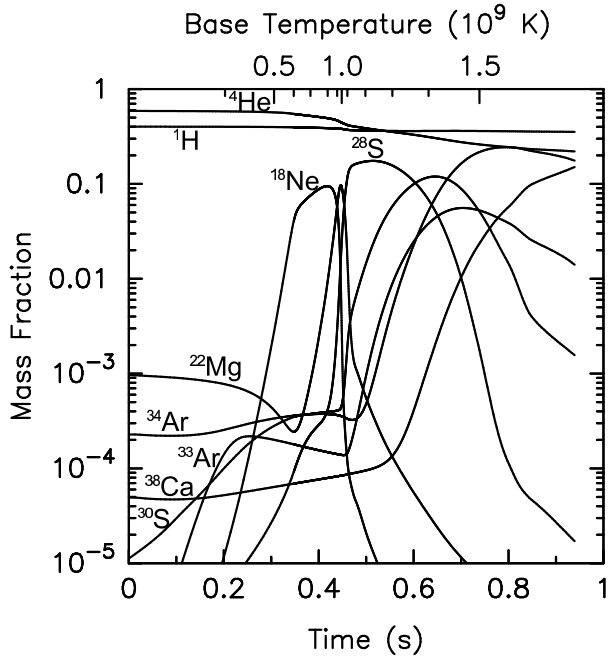}
\caption{ Mass fraction of various isotopes as a function of time
  ({\it bottom axis}) and base temperature ({\it top axis}) for model
  HHe0.1. Only those isotopes whose peak mass fraction exceeds 0.05 are
  shown.  \label{fig:AbundFlow2}}
\end{figure}

Figure~\ref{fig:Flow} shows a typical reaction flow during pure He
burning and Figure~\ref{fig:AbundFlow1} shows the composition as a
function of time and base temperature $T_b$. After a slow rise from
0.2 to 0.4~GK temperatures rise quickly, typically within 10-100~ms
from 0.4~GK to peak temperatures set by $y_b$ of 1.8~GK. The
reaction flow is mainly characterized by a sequence of
$\alpha$-captures into the $^{36}$Ar-$^{44}$Ti region.  However,
reactions off the $\alpha$-chain also play an important role.  For
example, beyond Mg, ($\alpha$,p)-(p,$\gamma$) sequences carry a
substantial part of the flow and in fact dominate in some cases, such
as for the flow from $^{36}$Ar to $^{40}$Ca, or the flow beyond
$^{44}$Ti.  The importance of ($\alpha$,p)-(p,$\gamma$) links and
their inverses in He burning has been pointed out before
\citep{Hashimoto:83, Timmes:00}.  

The most important effect of off $\alpha$-chain reactions
however is the bypass of the $^{12}$C($\alpha$,$\gamma$) reaction, the
slowest reaction in the $\alpha$-chain up to $^{32}$S for temperatures
$> 1 \trm{ GK}$, by the much faster
$^{12}$C(p,$\gamma$)$^{13}$N($\alpha$,p)$^{16}$O reaction sequence.
The bypass requires protons as catalysts, which are produced by
($\alpha$,p) reactions on heavier $\alpha$-chain nuclei such as
$^{24}$Mg, $^{32}$S and $^{36}$Ar. Initially, He burning proceeds via
the 3$\alpha$ reaction, building up $^{12}$C with some weak reaction
flow through $^{12}$C($\alpha$,$\gamma$) and successive $\alpha$
captures. At $^{24}$Mg, $^{32}$S, and $^{36}$Ar, ($\alpha$,p)
branchings are significant and lead to the build up of a small amount
of protons. For example, in the pure He $\dot m / \dot m_{\rm Edd} =
0.1$ model the proton abundance reaches 3$\times 10^{-9}$ at a
temperature of about 1~GK, which is sufficient for the
$^{12}$C(p,$\gamma$)$^{13}$N($\alpha$,p)$^{16}$O bypass to take over
the reaction flow from $^{12}$C to $^{16}$O. At this point a runaway
effect sets in as any increase in flow beyond $^{12}$C increases the
production of protons, which further increases the efficiency of the
$^{12}$C(p,$\gamma$)$^{13}$N($\alpha$,p)$^{16}$O bypass.  Only 2 ms
later at a temperature of 1.2~GK the flow from $^{12}$C to $^{16}$O
exceeds the flow via the 3$\alpha$ reaction and $^{12}$C is quickly
depleted.  As a result, the
$^{12}$C(p,$\gamma$)$^{13}$N($\alpha$,p)$^{16}$O reaction sequence
completely dominates the reaction flow towards heavy elements during
the convective burning stage.  {\it This is a very robust effect, which is
independent of the initial metallicity, and will always occur during
He burning under high temperature conditions that lead to the
production of $^{22}$Mg or heavier nuclei.}

\begin{figure*}[t]
\vspace{-1.5cm}
\includegraphics[bb=120 65 499 801, angle=-90,scale=0.8]{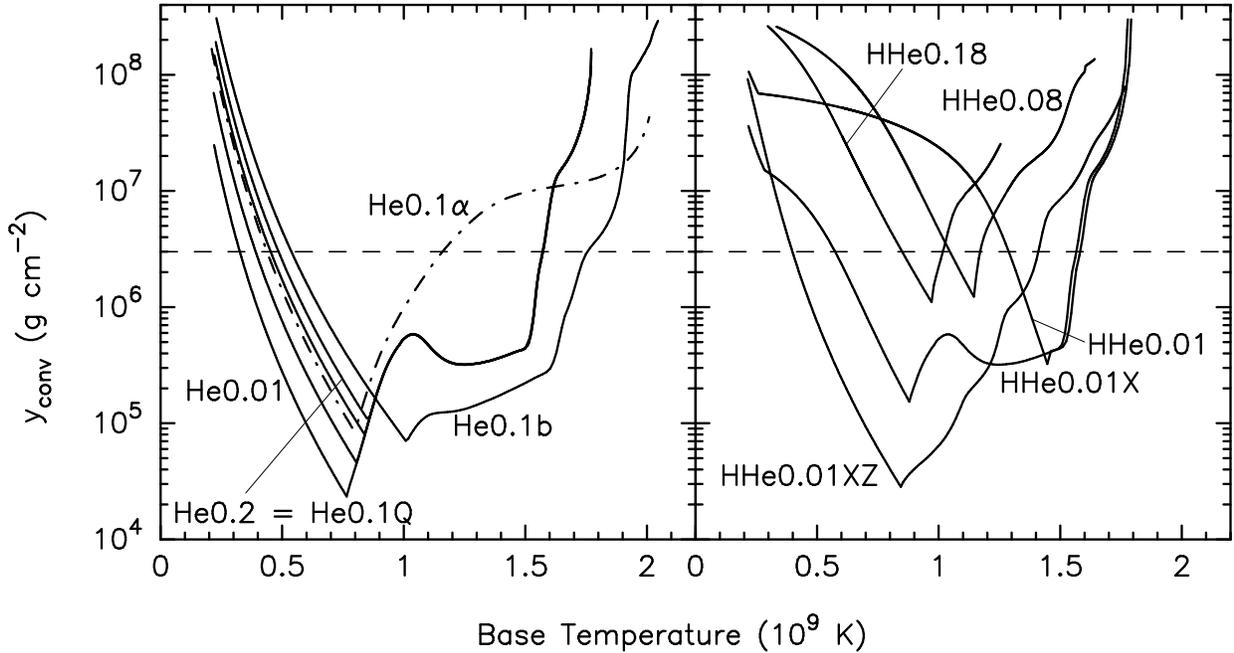}
\caption{Evolution of the top of the convective zone $y_c$ as a
  function of base temperature $T_b$ for burst models with pure helium
  accreted material ({\it left panel}) and mixed hydrogen-helium
  accreted material ({\it right panel}), as described in Table
  \ref{tab:1}. The two unlabeled helium models are, from bottom to
  top, He0.04 and He0.1. Model He0.1$\alpha$ ({\it dash-dot line}) was
  calculated using a simple $\alpha$-chain reaction network. The
  horizontal {\it dashed line} denotes the column depth where $y =
  0.01y_b$, corresponding approximately to the column depth $y_{\rm
    wind}$. \label{fig:ycTb}}
\end{figure*}

The rapid onset of the
$^{12}$C(p,$\gamma$)$^{13}$N($\alpha$,p)$^{16}$O bypass leads to a
sudden spike in energy production, which causes the drop in $t_{\rm
  gr}$ and $t_{\rm th}$ to $\sim \trm{ ms}$ time scales and the second
dip in $y_c$, as seen in Figures~\ref{fig:timescale} and
~\ref{fig:Ty}, respectively. As we show in \S~\ref{sec:results}, the
bypass occurs in all models burning helium-rich material regardless of
$\dot m$ (see Figure~\ref{fig:ycTb}), and is directly observable as a
rapid increase in the rise time of the burst light curve (see
Figure~\ref{fig:lightcurve}). For comparison, we also calculated the
pure He $\dot m / \dot m_{\rm Edd} = 0.1$ model with a simple
$\alpha$-chain network, the results of which are shown in
Figure~\ref{fig:ycTb}, \ref{fig:lightcurve}, and
\ref{fig:composition}. Another consequence of bypassing the
$^{12}$C($\alpha$,$\gamma$) reaction is the depletion of $^{12}$C and
the enhancement of the production of heavier $\alpha$-chain nuclei in
the final composition. As Figure~\ref{fig:composition} shows, a pure
He $\dot m / \dot m_{\rm Edd} = 0.1$ model with just an $\alpha$-chain
network would predict the main product to be $^{12}$C, while with the
full reaction network the main products are $^{28}$Si and $^{32}$S
with negligible amounts of carbon. In addition, the presence of
protons leads to proton capture reactions producing small amounts of
nuclei off the $\alpha$-chain. The very small amounts of nuclei around
$A=60$ are produced by proton capture of iron peak nuclei present in
the accreted material.

For pure He accretors, except for model He0.1b (see Table
\ref{tab:1}), the composition at the time the convective zone
has retreated back to the base is mainly doubly magic $^{40}$Ca
and unburned $^4$He, with some $^{36}$Ar and $^{44}$Ti. The most
abundant non $\alpha$-chain nucleus is $^{39}$K. He0.1b ignites
deeper, at a higher mass density, resulting in a higher peak
temperature. Burning therefore proceeds beyond $^{44}$Ti with a
sequence of alternating ($\alpha$,p) and (p,$\gamma$) reactions up to
$^{52}$Fe. Besides unburned He, the composition is then dominated by
$^{44}$Ti and $^{48}$Mn. For mixed H/He accretors that undergo pure He
burning at low accretion rates (e.g., models HHe0.01 and HHe0.01X) the
composition is very similar to the pure He accretion models except for
a somewhat lower $^{36}$Ar and higher $^{44}$Ti production. This
difference is due to the faster rise in temperature in the H/He
models, a consequence of the smaller convective extent in the presence
of hydrogen and thus a smaller convective mass to heat up for the same
energy input (see $t_{\rm gr}$ curve for He0.1 and HHe0.01 in
Figure~\ref{fig:timescale}).

Though some protons are produced during He burning, the presence of
large amounts of hydrogen at ignition in models such as HHe0.1
($X=0.4$) and HHe0.01XZ ($X=0.08$) leads to a drastic change in the
nuclear reaction sequence. Figure~\ref{fig:Flow} shows a typical
reaction flow and Figure~\ref{fig:AbundFlow2} shows the composition as
a function of time and base temperature $T_b$. The flow is governed by
the $\alpha$p-process, a sequence of ($\alpha$,p) and (p,$\gamma$)
reactions beginning at $^{14}$O \citep{Wallace:81}. For the larger
proton abundances in HHe0.1 several parallel side branches involving
more neutron deficient nuclei are established. The main product at the
time the convective zone returns to the base tends to be $^{38}$Ca
(37\% by mass) together with small amounts of $^{42}$Ti (7\% by mass).
Since He is abundant during the convective burning phase and since the
burning is largely confined to $Z \le 40$ nuclei, ($\alpha$,p)
reactions are always faster than $\beta$-decays. Thus, the rapid
proton capture process (rp-process) of \citet{Wallace:81}, a sequence
of proton captures and $\beta$-decays, does not yet play a role at
this early burst stage. It will take over at later times, producing
post-peak light curves like that seen in GS 1826-24 \citep{Galloway:04}.

\section{Results}
\label{sec:results}

\begin{figure}
\includegraphics[bb= 0 50 580 481, angle=-90,scale=0.42]{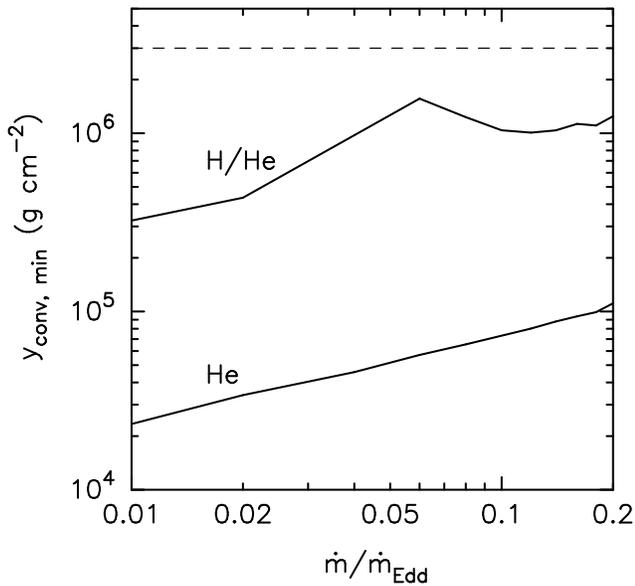}
\caption{Minimum column depth reached by the convective zone $y_{c, {\rm min}}$ as a function of  $\dot{m}$ in units of the Eddington rate. Results are shown for the pure helium accretion models He[$\dot m$] and the mixed hydrogen/helium accretion models HHe[$\dot m$]. The horizontal {\it dashed line} denotes the column depth where $y = 0.01y_b$, corresponding approximately to the column depth $y_{\rm wind}$. \label{fig:ycmdot}}
\end{figure}

The radiative wind generated by the super-Eddington luminosity of an RE burst will eject some ashes of burning if: (1) the fraction of the accreted mass ejected during the wind satisfies $\Delta M_{\rm w} / M_{\rm acc} = y_{\rm wind} / y_b > y_{c, {\rm min}} / y_b$, where $M_{\rm acc} \simeq 4 \pi R^2 y_b \approx 4 \times 10^{21} \trm{ g } (y_b / 3 \times 10^8 \trm{ g cm}^{-2})$, and (2) the wind is generated subsequent to the convective region reaching $y_c < y_{\rm wind}$. In \S~\ref{sec:dependence} we calculate $y_{c, \rm min}$ and demonstrate that condition (1) is satisfied for a wide-range of burst ignition models. In \S~\ref{sec:lightcurve} we calculate the light curve during burst rise and show that condition (2) is satisfied for those same models.  

We first note that since the gravitational binding energy at the surface is $\approx100$ times greater than the helium burning energy release per unit mass, at most $\approx 1\%$ of the atmosphere can be ejected by the wind.   To obtain an estimate of $\Delta M_{\rm w} / M_{\rm acc}$, suppose the surface luminosity as measured by an observer at infinity $L$ is super-Eddington so that the energy loss rate at the surface $\dot E \simeq L + \dot{M}_w c^2 / (1+z)$.  At the photospheric radius of the wind $r_{\rm ph} \gg R$ (see \S\ref{sec:lineswind}) the luminosity is very nearly the local Eddington value $L_{\rm ph} \simeq L_{\rm Edd}(r_{\rm ph}) = 4 \pi G M c (1+z_{\rm ph})/ \kappa \simeq 4 \pi G M c / \kappa$. Equating the energy loss rate at the surface to that at the photosphere yields the mass-loss rate due to the wind
\bea
\label{eq:dotmw}
\dot{M}_w   &\simeq&  \frac{L_{\rm Edd}}{c^2}\left(\frac{L}{L_{\rm Edd}} - 1\right) \left[1 - \frac{1}{1+z}\right]^{-1}\\ & \approx & 10^{18} \trm{ g s}^{-1} \left(\frac{0.2 \trm{ cm}^2 \trm{ g}^{-1}}{\kappa}\right) \left(\frac{L}{L_{\rm Edd}} - 1\right), \nonumber 
\eea
(\citealt{Paczynski:86}, see also \citealt{Wallace:82, Yahel:84, Joss:87, Nobili:94}). The burst duration $\Delta t \simeq M_{\rm acc} Q_{\rm nuc} / L$, where $Q_{\rm nuc} = 1.6 + 4.0 \langle X \rangle \trm{ MeV nucleon}^{-1}$ is the nuclear energy release and $ \langle X \rangle$ is the mass-weighted mean hydrogen fraction in the burning layer (C03), so that for a pure helium burst 
\bea
\frac{\Delta M_w}{M_{\rm acc}} & \simeq & \frac{\dot{M}_w \Delta t}{M_{\rm acc}} \simeq \left(1 -\frac{L_{\rm Edd}}{L}\right) \frac{Q_{\rm nuc}}{v_{\rm esc} c} \nonumber \\ 
& = & 0.003 \left(1 - \frac{L_{\rm Edd}}{L}\right).
\eea 

\subsection{Dependence of the convective extent on burst parameters}
\label{sec:dependence}

In Figure \ref{fig:ycTb} we show the time variation of the convective extent $y_c$ during a burst, with the progression marked by the temperature at the base of the burning layer, $T_b$, rather than by time.  We consider burst ignition models assuming both mixed hydrogen/helium accretion and pure helium accretion (see Table \ref{tab:1}); the latter models are similar to those obtained by C03 in fits to the burst properties of 4U 1820-30. We consider models over a range of accretion rate, ignition column depth, metallicity, and crustal heat flux. 

The evolution of $y_c$ is sensitive to the assumed ignition conditions and to the energy production from non-$\alpha$-capture reactions (see \S~\ref{sec:nucleosynthesis}). Ignition conditions that maximize the peak flux of a burst and minimize the entropy of the atmosphere during the fuel accumulation stage yield bursts with the smallest $y_{c, \rm min}$. This is because the larger the peak flux, the faster $T_b$ rises, and hence the shorter the growth time scale $t_{\rm gr}$ at a given $T_b$. The convective region must therefore reach out to lower column depths before $t_{\rm th}(y_c) = t_{\rm gr}$. As illustrated in Figure \ref{fig:ycmdot}, such ignition conditions are best satisfied at low $\dot m$. The reason is twofold. First, for $\dot{m} \la 0.04 \dot{m}_{\rm Edd}$ and solar abundance accreted material a pure helium layer develops at the base of the atmosphere and the peak luminosity is considerably greater than that of bursts with mixed hydrogen/helium burning layers.  Second, because the pre-burst flux $F_0 \propto \dot{m}$, the lower $\dot{m}$ the smaller the initial atmospheric entropy. 

For the models considered, the fraction of accreted mass lying above the convective region at its maximum extent is in the range $10^{-4}  \la \Delta M / M_{\rm acc} \la 10^{-2}$ making it likely that ashes get ejected.

\subsection{Light curve during burst rise and radiative winds}
\label{sec:lightcurve}

\begin{figure}
\includegraphics[bb= 0 50 580 481, angle=-90,scale=0.42]{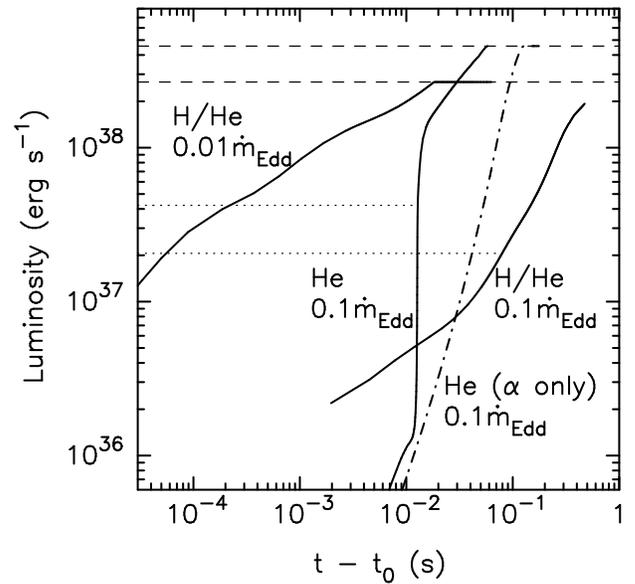}
\caption{Rising portion of the burst light curve for the same three models shown in Figure~\ref{fig:timescale} ({\it solid lines}). The {\it dash-dot line} is the light curve for the He0.1$\alpha$ model calculated using just an $\alpha$-chain reaction network rather than the full network.  The luminosity is plotted as a function of the offset time $t - t_0$ where $t_0$ corresponds to the time when the radiative region first begins to evolve away from its pre-burst state (i.e., when the equality $t_{\rm th} = t_{\rm gr}$ is first satisfied). The {\it dotted lines} denote the constant accretion luminosity for the models and the {\it dashed lines} denote the Eddington luminosity at the photosphere. \label{fig:lightcurve}}
\end{figure}

\begin{figure}
\includegraphics[bb= 0 50 580 481, angle=-90,scale=0.42]{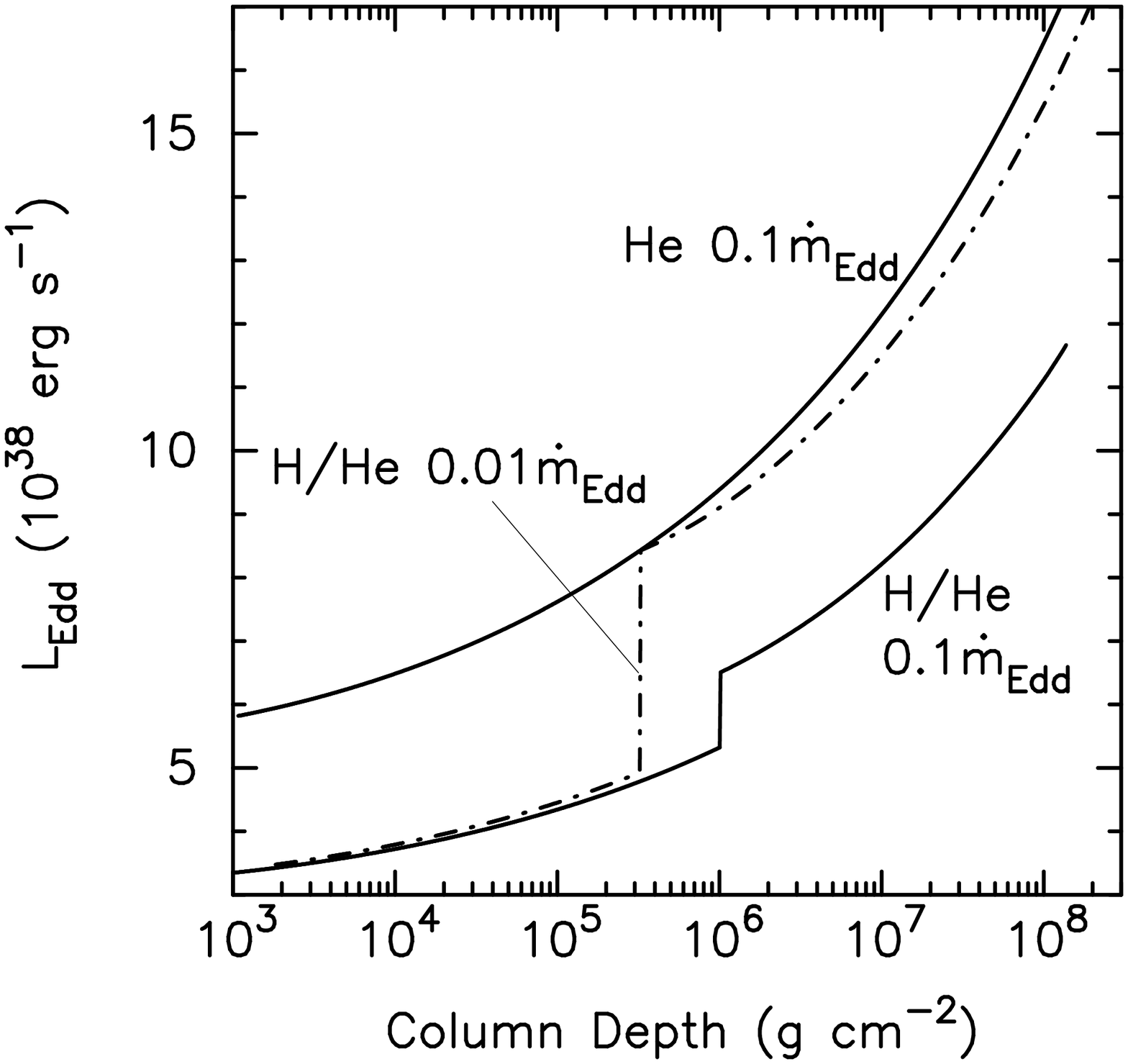}
\caption{ The local Eddington luminosity $L_{\rm Edd}$ as a function
  of column depth at the end of the calculation, when the convective
  zone has fully receded back to the base. The discontinuity at
  $y_{c, \rm min}$ for the H/He models is due to a discontinuity in
  hydrogen abundance between burnt and unburnt material.  \label{fig:Ledd}}
\end{figure}

We determine the rising portion of the burst light curve by calculating the radiative flux loss at the photosphere $F_{\rm ph}$ as $T_b$ increases. To obtain $F_{\rm ph}$ we integrate equation (\ref{eq:heat}) inwards (assuming the radiative zero solution at the outer boundary), varying the flux at the top until the radiative solution intersects the convective solution at the column depth $y_c$. Initially, $t_{\rm th} > t_{\rm gr}$ and $F_{\rm ph} = 0$. Eventually $t_{\rm th} = t_{\rm gr}$ and $F_{\rm ph}$ rises as the radiative region heats up due to the rise in $F_{\rm loss}(y_c)$.  In Figure \ref{fig:lightcurve} we show the rising portion of the light curve for the same models as Figure \ref{fig:timescale} and also for a pure helium model using only the $\alpha$-chain network. We plot the luminosity $L = 4 \pi R^2 F_{\rm ph}$ as a function of $t - t_0$, where $t_0$ corresponds to the time when the equality $t_{\rm th} = t_{\rm gr}$ is first satisfied. 

The rise time at a given time $t$ is set by the thermal time at the convective-radiative interface $t_{\rm th}(y_c[t])$. As the differences between the He0.1 and He0.1$\alpha$ models illustrate, the shape and rise time of the light curves are sensitive to the energy generation rate. The sharp sub-ms rise in models HHe0.01 and He0.1 occurs at the onset of the $^{12}$C(p,$\gamma$)$^{13}$N($\alpha$,p)$^{16}$O bypass (\S~\ref{sec:nucleosynthesis}). The bypass leads to a spike in energy production which in turn drives the convective zone to lower pressures. Since $T(y_c) = T_c$ is very high by this time, $t_{\rm th} = C_p T_c y_c / F_{\rm loss}(y_c) = 3 \kappa C_p y_c^2 / 4 a c \lambda n T_c^3$ becomes very small (see Figure~\ref{fig:timescale}). In the He0.1 model, the light curve initially rises relatively slowly ($\ga 10 \trm{ ms}$) and then, once the temperature is high enough for the bypass to take over, it sharply rises. This is distinct from the HHe0.01 model which rises sharply at the outset because $t = t_0$ nearly coincides with the onset of the bypass.

Depending on the ignition conditions, the base temperature can get sufficiently high that the flux becomes super-Eddington. In Figure~\ref{fig:Ledd} we show the local Eddington luminosity $L_{\rm Edd}(y)$ as a function of column depth for the models HHe0.01, HHe0.1, and He0.1 at the end of the calculation, when the atmosphere is completely radiative (i.e., $y_c = y_b$). The increase in $L_{\rm Edd}(y)$ with depth is the result of the decrease in $\kappa_{\rm es}$ with temperature due to relativistic corrections. The discontinuity in $L_{\rm Edd}(y)$ for the mixed H/He models occurs at $y_{c, \rm min}$ and is due to the discontinuity in hydrogen abundance between the burnt and unburnt material. Such a discontinuity could explain the possible bimodal distribution of the observed burst peak luminosities (see \citealt{Kuulkers:03} for a recent discussion), as first pointed out by \citet{Sugimoto:84} and investigated in greater detail by \citet{Ebisuzaki:88}. They proposed that the ejection of the hydrogen-rich envelope during radius expansion burst occasionally exposes the underlying helium-rich region and results in the Eddington limit transitioning from that of hydrogen-rich material to that of hydrogen-poor material. 

\begin{figure}
\includegraphics[bb= 0 50 580 481, angle=-90,scale=0.42]{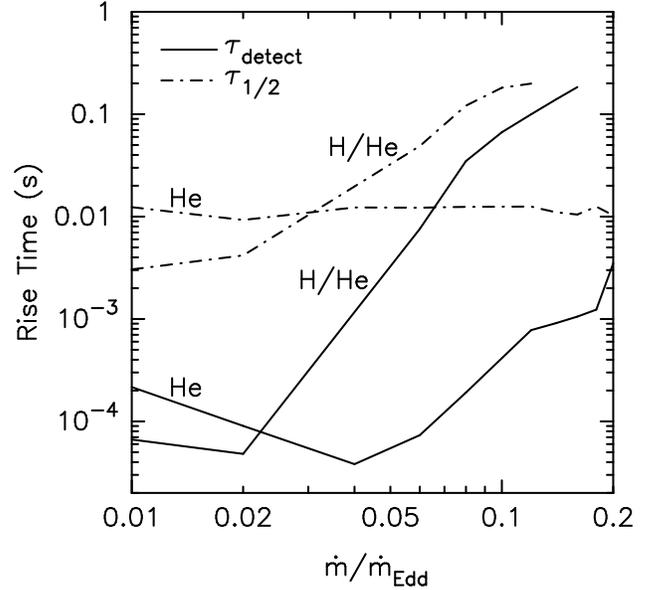}
\caption{Rise time of the burst light curve as a function of $\dot{m}$ in units of the Eddington rate for the same models as Figure~\ref{fig:ycmdot}. Two sets of rise times are computed for each model: $\tau_{\rm detect}$ ({\it solid lines}) is the instantaneous rise time $dt / d\ln L$ when the luminosity first exceeds the accretion luminosity, corresponding to when an observer can first detect the burst;  $\tau_{\rm 1/2}$ ({\it dash-dot lines}) is the time it takes the burst to rise from 10\% to 50\% of its peak luminosity. \label{fig:trisemdot}}
\end{figure}

We find that the peak luminosity exceeds $L_{\rm Edd}(y)$ for $\dot m \la 0.1 \dot{m}_{\rm Edd}$ if the accreted material has solar abundance ($\dot m$ can be higher if the material is helium-rich) and that the Eddington value is first exceeded near the top of the atmosphere. Such systems are expected to develop a radiative wind. Eventually, the flux loss exceeds the flux from nuclear burning and the atmosphere cools, though we do not calculate this portion of the light curve. Note that since the convective zone reaches a minimum pressure at $t \simeq t_0$, bursts that become super-Eddington achieve $y_c < y_{\rm wind}$ well before the wind is generated. 

In Figure \ref{fig:trisemdot} we show the dependence of the rise time on $\dot m$ for the pure He models and the mixed H/He models. We calculate rise times during the early portion of the rise, when the surface flux just exceeds the accretion flux ($\tau_{\rm detect}$), and during the later portion of the rise, taken to be the time for the burst to rise from 10\% to 50\% of its peak luminosity ($\tau_{1/2}$). Since  $t_{\rm th}(y_{c, {\rm min}})$ sets the initial rise time, models with smaller values of $y_{c, \rm min}^2 / T_{c, \rm min}^3$ tend to have smaller $\tau_{\rm detect}$, which, to a first approximation, corresponds to models with lower $\dot m$. For helium accretors, $\tau_{1/2}$ is nearly independent of $\dot m$ because the convective zone evolution at late times is very similar among models that differ only in $\dot m$ (see Figure~\ref{fig:ycTb}). By contrast, for accretion of hydrogen-rich material, $\tau_{1/2}$ increases by a factor of $\approx 50$ between $\dot{m}/\dot{m}_{\rm Edd} = 0.02$ and 0.1.  In their multizone numerical simulations of X-ray bursts, \citet{Woosley:04} consider models that accrete mixed hydrogen/helium with solar abundance at $\dot{m}/\dot{m}_{\rm Edd} = 0.02$ and 0.1. They follow several sequences of bursts for each model and obtain values of $\tau_{1/2}$ in the range $0.51\times10^{-3}-32.1\times10^{-3} \trm{ s}$ for $\dot{m}/\dot{m}_{\rm Edd} = 0.02$ and $0.51-0.66 \trm{ s}$ for $\dot{m}/\dot{m}_{\rm Edd} = 0.1$, both in good agreement with our estimates. Our rise times are also in broad agreement with the relevant models of earlier numerical simulations (e.g., \citealt{Taam:81, Ayasli:82, Wallace:82}).

\section{Detecting the Nuclear Burning Ashes}
\label{sec:detecting} 
 
In \S~\ref{sec:results} we showed that the convective zone reaches sufficiently low pressures during RE bursts that $y_{c, {\rm min}} \ll y_{\rm wind}$ over a broad range of burst parameters. Thus, ashes of nuclear burning can be ejected by the radiative wind of RE bursts. Furthermore, when the photosphere settles back down to the NS surface following the RE phase, it is laced with heavy-element ashes. In this section we describe the composition of the ejected and exposed ashes (\S~\ref{sec:composition}) and address whether they can be detected. In \S~\ref{sec:ejection} we discuss the possibility that some of the p-nuclei found in the solar system owe their origin to ash ejection during RE bursts. In \S\S~\ref{sec:lineswind} and \ref{sec:linessurface} we determine the expected strength of photoionization edge features from ashes ejected in the wind and those exposed at the NS surface.

\subsection{Composition of ejected and exposed ashes}
\label{sec:composition}
 
\begin{figure*}[!ht]
\includegraphics[bb= 187 60 500 281, angle=-90,scale=0.8]{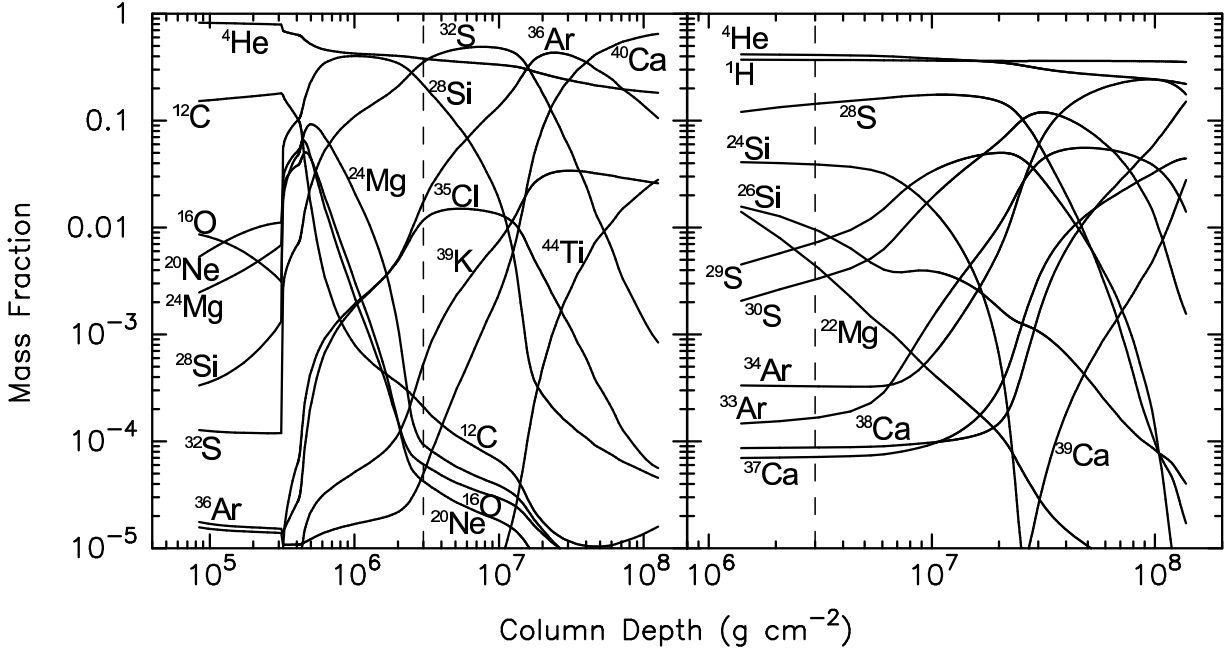}
\caption{Ash composition as a function of column depth after the convective zone has completely receded (i.e., $y_c = y_b$). Results are shown for models He0.1 ({\it left panel}) and HHe0.1 ({\it right panel}). Only those isotopes whose peak mass fraction exceeds 0.01 somewhere in the range $y_{c, \rm min} < y < y_b$ are shown. The vertical {\it dashed line} denotes $y =  0.01 y_b \simeq y_{\rm wind}$. \label{fig:abundyc1}}
\end{figure*}

\begin{figure*}[!ht]
\includegraphics[bb= 60 110 400 441, angle=0,scale=0.8]{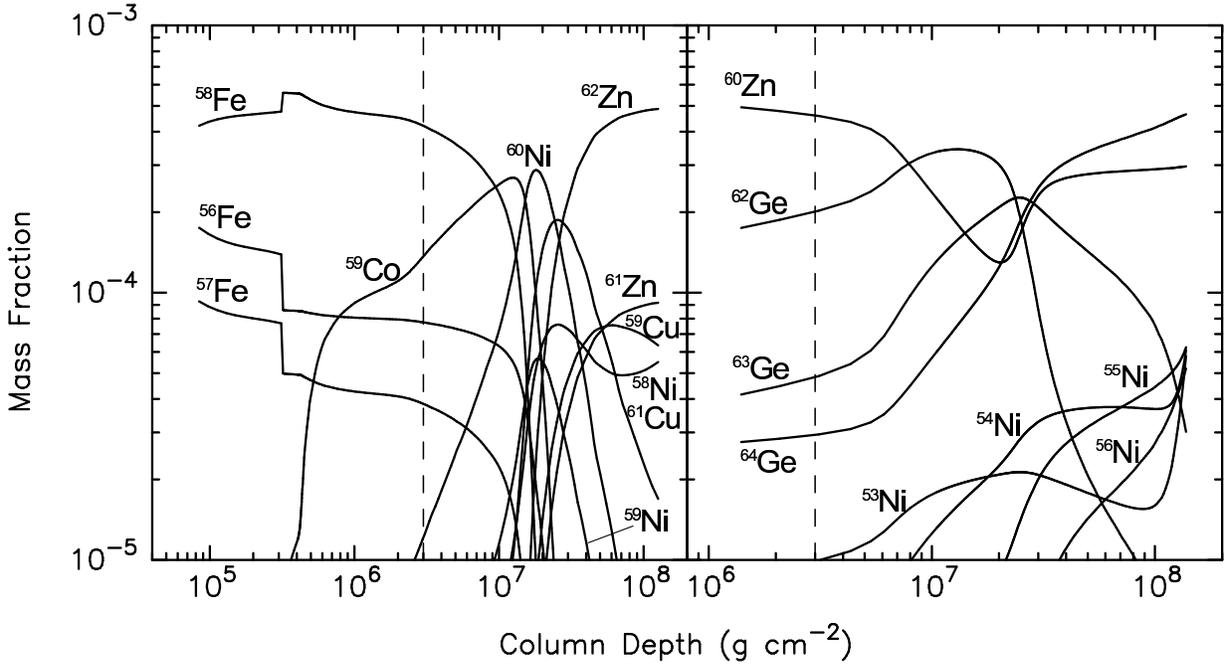}
\caption{Same as Figure~\ref{fig:abundyc1} except only those isotopes with $A > 50$ whose peak mass fraction exceeds $5\times10^{-5}$ somewhere in the range $y_{c, \rm min} < y < y_b$ are shown.\label{fig:abundyc2}}
\end{figure*}

\begin{figure}
\includegraphics[bb= 0 50 580 481, angle=-90,scale=0.42]{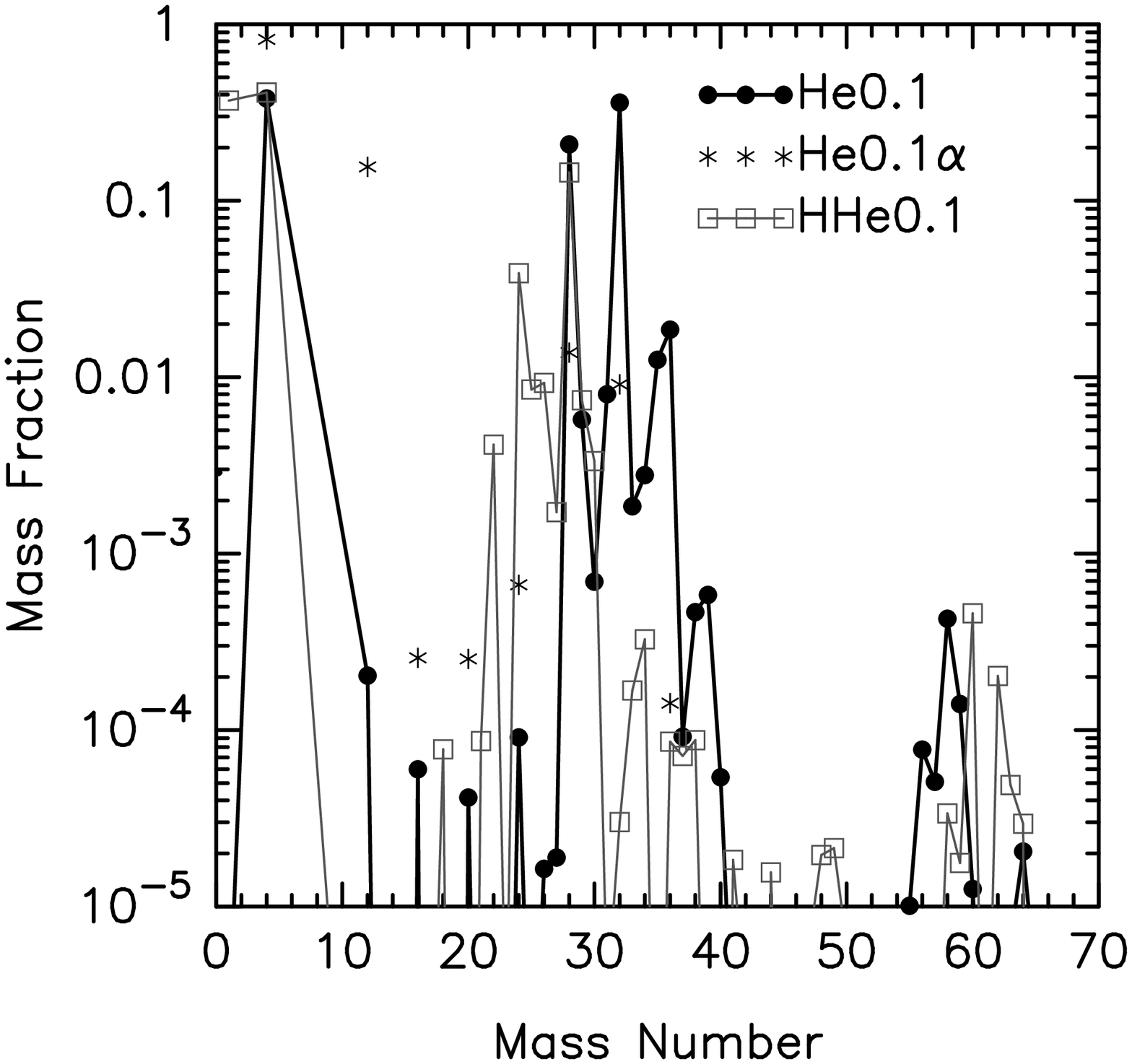}
\caption{Composition of material processed during a burst at the time the convective zone is located at $y_c = 0.01 y_b \simeq y_{\rm wind}$ and is receding to higher pressures. Results are shown for models HHe0.1, He0.1, and He0.1$\alpha$.  \label{fig:composition}}
\end{figure}
 
Just prior to the onset of the wind, the convective zone has receded to the base and the atmosphere has a stratified compositional structure. Throughout the region $y < y_{c, {\rm min}}$, the composition is that of the unprocessed accreted material. As we show in Figures \ref{fig:abundyc1} and \ref{fig:abundyc2}, for $y >  y_{c, {\rm min}}$ the composition is determined by the burning stage at the moment $y_c(t) = y$ during the convective zone's retreat to the base.  Results are shown for models He0.1 and HHe0.1. The discontinuity at $y = 3 \times 10^5 \trm{ g cm}^{-2}$ in model He0.1 is due to the rapid nucleosynthesis that occurs once the $^{12}$C(p,$\gamma$)$^{13}$N($\alpha$,p)$^{16}$O bypass takes over the reaction flow from $^{12}$C to $^{16}$O (\S~\ref{sec:nucleosynthesis}). The composition at the base of the wind is approximately that at $y = 0.01 y_b = 3 \times 10^6 \trm{ g cm}^{-2}$. In Figure \ref{fig:composition} we show the ash composition at $y / y_b = 0.01 \simeq \Delta M_{\rm w} / M_{\rm acc}$ for models He0.1, He0.1$\alpha$, and HHe0.1. While helium comprises the largest fraction of the mass in all three models, the models' overall abundance distributions differ significantly from one another due to the impact of the $^{12}$C(p,$\gamma$)$^{13}$N($\alpha$,p)$^{16}$O bypass and differences in the initial composition at the base of the burning layer. Some of the proton-rich isotopes shown in the above figures have half-lifes comparable to the duration of the atmosphere's convective phase. Though we do not show the decay products in these figures, we do account for such decays in our calculation of photoionization edges below. 

\subsection{Ejection of p-nuclei}
\label{sec:ejection}
  
In their numerical simulations of X-ray bursts, \citet{Woosley:04} find that bursts ignite in the ashes of previous bursts. If so, these endpoint ashes, which are heavier than those processed during the ongoing burst at the time $y_c / y_b = 0.01$, are mixed throughout the convective region (we do not show such ashes in Figures \ref{fig:abundyc1}, \ref{fig:abundyc2}, or \ref{fig:composition}). Using a reaction network that extends up to Xe, \citet{Schatz:01} find that in models where solar abundance material is accreted at $\dot m = 0.3 \dot{m}_{\rm Edd}$ (i.e., burning in a hydrogen-rich environment), the endpoint of rp process burning is a closed SnSbTe cycle that naturally limits rp process nucleosynthesis to light p nuclei. They find overproduction factors (relative to solar abundance) of $\sim10^{8}$ for the p nuclei $^{92}$Mo and $^{96}$Ru and $\sim10^9$ for the p nucleus $^{98}$Ru (see also \citealt{Schatz:98}). Standard p-process scenarios are unable to adequately explain the observed solar system abundances of these p-nuclei (for a review, see \citealt{Wallerstein:97}).  

Whether ash ejection during RE bursts can account for the observed solar system abundances depends on the amount of p-process material ejected into the interstellar medium by an RE burst and the event rate of such bursts over the Galaxy lifetime. In order to produce large amounts of p-nuclei, the burning layer must be hydrogen-rich ($\dot m \approx 0.3 \dot{m}_{\rm Edd}$) while RE bursts require helium-rich burning layers and thus $\dot m \la 0.05 \dot{m}_{\rm Edd}$. However, this does not preclude RE bursts from ejecting p-nuclei, as accretion rates in bursting low-mass X-ray binaries are observed to vary by factors of a few, with individual systems undergoing transitions from hydrogen-rich to helium-rich burning over year time scales  (as evidenced by variations in burst duration and peak fluxes; see e.g., \citealt{Cornelisse:03}). Another limitation to how much p-nuclei an RE burst can eject is that the p-nuclei tend to be produced at the bottom of the burning zone where the hydrogen is consumed. Thus the ignition and base of the burning zone may lie well above the p-rich ashes of a preceding burst. Unless the convective zone dredges up ashes from deep below, the amount of p-nuclei mixed into the convective zone may be small.

To determine the fractional amount $\eta$ of ashes that must be ejected in order to account for the observed solar system abundance of p-nuclei, assume a p-nuclei overproduction factor $\xi = 10^9$ and a galaxy disk mass $M_{\rm disk} = 4 \times 10^{10}$ \citep{Klypin:02}. Currently, there are $\approx 10$ active X-ray burst systems at $\dot M \sim 10^{-9} M_\odot \trm{ yr}^{-1}$ and $\approx 100$ at $\dot M \sim 10^{-10} M_\odot \trm{ yr}^{-1}$ \citep{Lewin:95}. If we assume this is representative of the population count over the galaxy lifetime, then the total amount of mass accreted by all RE burst systems over $10^{10} \trm{ yr}$ is $M_{\rm acc, tot} \sim 100 M_\odot$. A fraction $\eta = 0.4  (M_{\rm disk} / 4\times10^{10} M_\odot)(\xi/10^9)^{-1}(M_{\rm acc, tot}/100 M_\odot)^{-1}$ of all accreted material must therefore be ejected. Thus, $\eta$ is a factor of $\sim 100$ too high given that only $\approx 1\%$ of all accreted matter is ejected. The discrepancy can be overcome if, for example, the Galactic distribution of p-nuclei is inhomogeneous (i.e., the solar abundance is higher than the Galactic mean by a factor of $\sim 100$) or the population of bursting systems was much larger at earlier times in our galaxy. 

\subsection{Spectral Edges in Wind Outflow}
\label{sec:lineswind}
                                 
During the RE phase, an optically thick, transonic, radiation-driven wind forms. The sonic point of the wind lies $10-100 \trm{ km}$ above the NS surface and the photosphere, defined as the location where the effective optical depth $\tau_\ast \equiv \kappa \rho r$ is near unity, is a factor of $\sim 10$ farther out \citep{Paczynski:86, Joss:87, Nobili:94}. Once matter reaches the sonic point it is essentially unbound from the NS, and is ejected to infinity. 

The ejected matter is at sufficiently low temperature during the RE phase that some heavy elements bind with one or more electrons. The resulting column density of hydrogen-like ions above the photosphere is thus $N_{\rm wind} \sim f(A, Z) \zeta(T, \rho) N_e \simeq 10^{20} \trm{ cm}^{-2} (\zeta f / 10^{-4})$, where $f(A, Z)$ is the abundance by number of element $Z$ with mass number $A$, $\zeta(T, \rho)$ is the fraction in the hydrogen-like state at a given temperature and density from Saha equilibrium, and $N_e \approx \sigma_{\rm Th}^{-1}$ is the electron column density.  This corresponds to an optical depth to a photoionization edge of $\tau \approx N_{\rm wind} \sigma_{\rm bf} \sim 1$, where the bound-free cross section $\sigma_{\rm bf}(E) = 6.32 \times 10^{-18} \trm{ cm}^2 (E_e / E)^3 Z^{-2}$ and $E_e \simeq 13.6 Z^2 \trm{ eV}$ is the edge energy \citep{Rybicki:79}. The effective temperature is low enough at the photosphere that even for metals at solar abundances $\tau \sim 1$. That the ejected ashes have abundances much larger than solar improves the likelihood of detection.

We now calculate $N_{\rm wind}$ more exactly and determine the resulting equivalent width (EW) of the photoionization edge. The column density of hydrogen-like ions for a given species is
\beq
N_{\rm wind} = \frac{f(A, Z)}{m_p} \int_{r_{\rm ph}}^\infty \zeta\left[T(r),\rho(r)\right]\rho(r) \, dr.
\eeq
The recombinations are nearly instantaneous on the time scale of the wind flow, i.e., $t_{\rm rec} \ll t_{\rm flow} \sim r/v(r)  \approx 0.1 \trm{ s}$, where  $v(r)$ is the fluid velocity at $r$ and $t_{\rm rec} \equiv 1 / n_e \langle \sigma_{\rm fb} v \rangle \approx 0.01 Z^{-4} \rho^{-1}_{-6} T_7 ^{-3/2} B(Z, T) \trm{ s}$, with $\rho_{-6} = \rho /10^{-6} \trm{ g cm}^{-3}$, $T_7 = T/10^7 \trm{ K}$, $B(Z,T) = \exp(x) \int_x^\infty \exp(-t) \, d\ln t$, $x\equiv E_e/kT$, and $\sigma_{\rm fb}$ related to  $\sigma_{\rm bf}$ through the Milne relation \citep{Rybicki:79}. We use the results of \citet{Paczynski:86}, who calculate general relativistic models of radiation-driven winds from NSs, to obtain values for $r_{\rm ph}$,  $T(r_{\rm ph})$, and $\rho(r_{\rm ph})$  as a function of the mass outflow rate $\dot M_{\rm w}$. For example, at $\dot M_{\rm w} = 10^{18} \trm{ g s}^{-1}$, $r_{\rm ph} = 3 \times 10^7 \trm{ cm}$, $T_{\rm ph} = 5\times10^6 \trm{ K}$, and $\rho_{\rm ph} = 5 \times 10^{-7} \trm{ g cm}^{-3}$.  The wind duration is longer than $t_{\rm flow}$ so we assume that $\dot M_{\rm w} \simeq 4 \pi r^2 \rho v$ (see e.g., \citealt{Joss:87}).  \citet{Paczynski:86} show that the velocity is nearly constant beyond the photosphere, i.e., $v(r > r_{\rm ph}) \simeq v_{\rm ph} \sim 10^8 \trm{ cm s}^{-1}$, so that $\rho(r>r_{\rm ph}) \approx \dot M_{\rm w} / 4 \pi r^2 v_{\rm ph}$. 

The gas above the photosphere is Compton-heated by the hot photons originating in the photosphere. Since the Compton heating time scale $t_{\rm c} \sim kT m_e c^2 / E_\gamma F \sigma_{\rm Th} \sim 10^{-7} \trm{ s} \ll t_{\rm flow}$, where $E_\gamma$ is the photon energy and $F$ the flux, the gas temperature is nearly constant out to radii well above the photosphere \citep{Joss:87, Nobili:94}. For $r \gg r_{\rm ph}$, $t_{\rm c} \ga t_{\rm flow}$ and the gas cools adiabatically, though the density in this region is so low that in calculating $N_{\rm wind}$ we assume $T(r>r_{\rm ph}) = T_{\rm ph}$. For the more abundant species shown in Figure \ref{fig:composition}, we obtain column densities in the range $N_{\rm wind} \sim 10^{16}-10^{21} \trm{ cm}^{-2}$.

\begin{figure}
\includegraphics[bb= 0 50 580 481, angle=-90,scale=0.42]{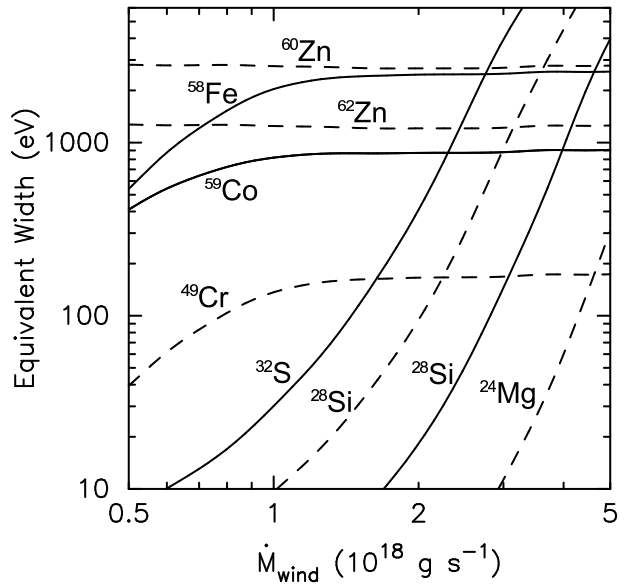}
\caption{Equivalent width of the photoionization edge due to ashes ejected in a wind as a function of the mass outflow rate for the models He0.1 ({\it solid lines}) and HHe0.1 ({\it dashed lines}). \label{fig:ewwind}}
\end{figure}

To determine the EW of the photoionization edge we integrate over the optical depth above the edge, $\tau_e = N_{\rm wind} \sigma_{\rm bf}(E)$. Thus, assuming an effectively cold atmosphere, $\trm{EW}_e = \int \lbrace1-\exp[-N_{\rm wind} \sigma_{\rm bf}(E)]\rbrace \, dE$ \citep{Bildsten:03}. In Figure \ref{fig:ewwind} we show EW$_e$ as a function of $\dot M_{\rm w}$ for some of the ejected ashes of models He0.1 and HHe0.1. In calculating the EWs we accounted for the decay of those isotopes whose half-lifes are shorter than $1 \trm{ s}$, and therefore shorter than the wind duration. Thus, $^{24}$Si$\rightarrow$$^{24}$Mg,  $^{28}$S$\rightarrow$$^{28}$Si,  $^{49}$Fe$\rightarrow$$^{49}$Cr, and $^{62}$Ge$\rightarrow$$^{62}$Zn, all from model HHe0.1. As $\dot M_{\rm w}$ increases, $T_{\rm ph}$ decreases and the hydrogen-like fraction $\zeta$ increases. For the heavier ashes (e.g.,  $^{60}$Zn and $^{62}$Zn)  $\zeta \sim 1$ even at low $\dot M_{\rm w}$ so that $N_{\rm wind}(Z\sim30)$ and EW$_e(Z\sim30)$ are essentially set by $f(A, Z)$, as noted above.  The computed values of EW$_e$ are within the range accessible by current X-ray telescopes. 

Several observational studies report the possible presence of spectral lines and edges during the RE phase of bursts. Indications of such features were found in RXTE/PCA observations of GX 17+2 \citep{Kuulkers:02}, 4U 1722-30, 4U 2129+12, XB 1745-25 \citep{Kuulkers:03}, and 4U 1820-30 \citep{Strohmayer:02, Kuulkers:03}, in Ginga/LAC observations of 4U 2129+12 \citep{vanParadijs:90}, in EXOSAT observations of EXO 1747-21 \citep{Magnier:89}, and Tenma observations of 4U 1636-53 \citep{Waki:84}. In general, these studies report significant residuals in black-body spectral fits that cannot be explained by spectral hardening alone. Higher resolution observations with Chandra and XMM-Newton are clearly needed.  

\subsection{Spectral Edges from NS surface}
\label{sec:linessurface}

Following the burst peak, the atmosphere cools, the flux becomes sub-Eddington, and the wind turns off. The photosphere, still laced with heavy-element ashes, settles back down to the NS surface. As in the case of spectral edges in the wind outflow, whether these surface ashes can be detected depends on the column density, $N_{\rm surf}$, of ashes that are not fully ionized. Here $N_{\rm surf} \simeq f(A,Z) \zeta(T, \rho) y_{\rm ph} / m_p$, where $y_{\rm ph} \approx 1 \trm{ g cm}^{-2}$.  As the surface temperature $T$ decreases, $N_{\rm surf}$ increases, although edges cannot be detected once $L = 4 \pi R^2 \sigma T^4 \la L_{\rm acc}$, where $L_{\rm acc}$ is the accretion luminosity. The decay time scale of RE bursts (i.e., the time during which $L_{\rm acc} < L < L_{\rm Edd}$) is typically around $\sim 10\trm{ s}$. The downward drift speed of a nucleus with $Z\sim30$ in a pure H atmosphere is $v\approx 1 \trm{ cm s}^{-1} T_7^{3/2}/\rho$ so that it takes $t_s \approx 1 \trm{ s} (y / 1 \trm{ g cm}^{-2})/T_7^{3/2}$ for such a nucleus to fall a column depth $y$ \citep{Bildsten:92,Bildsten:03}. Thus, the residence time of ashes in the photosphere also limits the detectability of edges, as will high NS rotation rates \citep{Ozel:03, Chang:05}.

\begin{figure}
\includegraphics[bb= 0 50 580 481, angle=-90,scale=0.42]{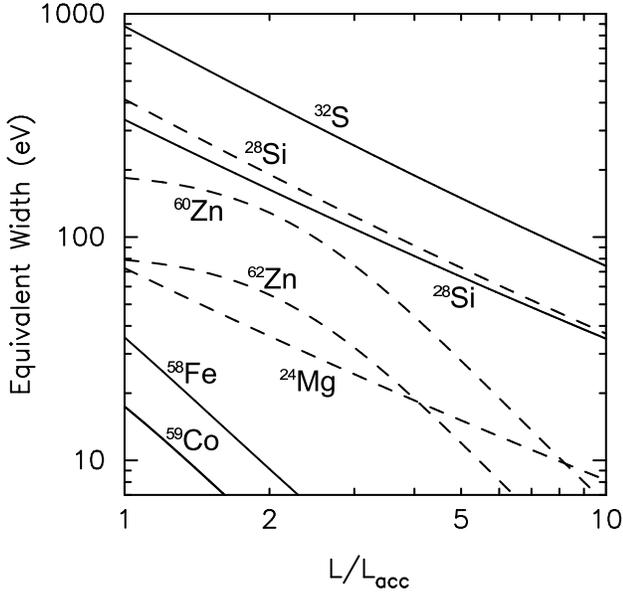}
\caption{Equivalent width of the photoionization edge due to ashes residing in the photosphere at the NS surface for models He0.1 ({\it solid lines}) and HHe0.1 ({\it dashed lines}). The EW is plotted as a function of the ratio of surface luminosity to accretion luminosity for a cooling NS atmosphere following the RE phase. Detecting such edges might allow for a measurement of the NS gravitational redshift. \label{fig:ewsurface}}
\end{figure}

In Figure \ref{fig:ewsurface} we plot EW$_e = \int \lbrace1-\exp[-N_{\rm surf} \sigma_{\rm bf}(E)]\rbrace \, dE$ as a function of $L/L_{\rm acc}$ for the same models as Figure \ref{fig:ewwind}. For reference, $L_{\rm Edd} / L_{\rm acc} \simeq 10$ and 30 for models He0.1 and HHe0.1, respectively.  The values of EW$_e$ are again within the range accessible by current X-ray telescopes and a measurement of the NS gravitational redshift may be possible. 

\section{Summary and Conclusions}
\label{sec:summary} 

We have shown that during a radius expansion Type I X-ray burst the ashes of nuclear burning can be ejected by the burst's radiative wind. Specifically, we solved for the evolution of the atmosphere's thermal structure and found that in systems accreting pure helium, such as 4U 1820-30, and in systems accreting mixed hydrogen and helium at $\dot m \la 0.05 \dot m_{\rm Edd}$, the convective zone reaches sufficiently low pressures that it lies within the wind region located at pressures $\la 1\%$ that at the base. Hence ashes of burning, mixed throughout the convective zone, can be amongst the wind ejecta. Previous studies have also found that the convective zone can reach $\la 1\%$ the base pressure for low $\dot m$ bursts \citep{Joss:78, Taam:81,Ayasli:82, Hanawa:82,Hanawa:84, Woosley:04}. However, they focused on numerical simulations of bursts and typically did not resolve the low-pressure zones convection reaches. Furthermore, such studies only explored a limited range of burst parameter space. Our analytic treatment enabled a survey of the dependence of the minimum pressure reached by the convective zone on a variety of burst parameters such as $\dot m$, the composition of accreted matter, and the pre-burst thermal state of the atmosphere. We have compared some of the results of our analysis, such as the burst rise times, to those of numerical simulations and found good agreement.

To calculate the nuclear energy generation rate and ash composition we used an updated full reaction network that contains 394 nuclei and all relevant proton and $\alpha$ particle induced reactions and the corresponding inverse processes. We have found that in models where burning initiates in a pure helium layer (i.e., those systems accreting mixed hydrogen and helium at $\dot m \la 0.04 \dot m_{\rm Edd}$ or systems accreting pure helium), protons are produced by $(\alpha, p)$ reactions on heavier $\alpha$-chain nuclei such as $^{24}$Mg, $^{32}$S, and $^{36}$Ar. The proton abundances achieved are high enough to enable the bypass of the relatively slow $^{12}$C$(\alpha, \gamma)$ reaction by the much faster $^{12}$C$(p, \gamma)^{13}$N$(\alpha, p)^{16}$O reaction sequence. The bypass leads to a burst in energy production that pushes the convective zone to low pressures and results in a rapid, $\sim 10^{-3} \trm{ s}$, rise in the burst light curve (see Figure~\ref{fig:trisemdot}). The bypass also enhances the production of heavier $\alpha$-chain nuclei, resulting in heavier ejected ashes.  

For specific burst models we determined the composition of ejected ashes and calculated the expected column density of hydrogen-like nuclei using models of relativistic radiation-driven winds. We then computed the EW of the photoionization edge for the more abundant hydrogen-like nuclei in the wind. We carried out a similar procedure to determine the EW for those ashes that remain bound to the NS and thus reside in the photosphere after it settles back down to the NS surface. We found EWs in the range $10 - 1000 \trm{ eV}$ (see Figures~\ref{fig:ewwind} and ~\ref{fig:ewsurface}) suggesting that the edges can be detected with current high resolution X-ray telescopes. Detecting them would directly probe the nuclear burning. Those edges formed at the surface may also provide a measurement of the NS gravitational redshift and thus help constrain the NS equation of state.

If bursts ignite in the ashes of previous bursts then some of these processed ashes, which are thought to contain large overabundances of p-nuclei relative to solar, are also ejected in the wind. We showed that at least $\sim 1\%$ of the p-nuclei observed in the solar system may originate in X-ray bursts. 

We did not account for the affect of a laterally spreading burning front during burst rise nor the influence of rotation, though these may significantly alter the convective structure and therefore the conditions under which ashes are ejected. Bursting systems have magnetic fields $B \la 10^9 \trm{ G}$, so the magnetic energy density, $B^2/8\pi$, is considerably smaller than even the minimum convective energy density  $\rho v_{\rm conv}^2 \simeq g y_{c, {\rm min}} \simeq 5\times10^{18} \trm{ erg cm}^{-3}$. Magnetic fields are therefore unlikely to significantly affect the convective evolution.

Although low $\dot{m}$ systems accreting a mix of hydrogen and helium yield X-ray bursts with convective zones that reach low pressures, their burst recurrence times are long and often irregular making them difficult targets to monitor given a narrow window of observing time. More promising are systems in which the neutron star accretes helium-rich material ($X_{\rm He} \approx 0.9 - 1$) from an evolved companion, such as a cold helium white dwarf. The binary 4U 1820-30 is thought to reside in such a system as evidenced by its ultracompact nature ($P_{\rm orb} = 11.4 \trm{ min}$; \citealt{Stella:87}) and the fast rise times, decay times, and $\alpha$-values observed during bursts (\citealt{Cumming:03}, and references therein).  The system also exhibits radius expansion bursts with fairly regular burst recurrence times of only a few hours (see e.g., \citealt{Cornelisse:03}, \citealt{Kuulkers:03}). 

Observations of four candidate ultracompact binaries have shown an unusual Ne/O abundance ratio in the absorption along the line of sight, with ratios several times the interstellar medium (ISM) value \citep{Juett:01, Juett:03}. Two of the systems, 4U 1543-624 and 4U 1850-087, have shown variations in the Ne/O ratio in follow-up observations \citep{Juett:03, Juett:05}, suggesting either variations in the ionization state of the Ne and O or variations in the intrinsic abundances. Both possibilities imply the absorption is due to material local to the binaries. 

One explanation for the unusual ratios is that the intrinsic abundance of Ne and O is the same as the ISM but the O is in a higher ionization state than Ne, leading to an apparent enhancement of the Ne/O ratio. Another possibility is that the degenerate donors in these ultracompact binaries have Ne/O abundances above the solar value. In particular, the donors may be the chemically fractionated cores of C-O-Ne or O-Ne-Mg white dwarfs that have previously crystallized \citep{Schulz:01, Juett:03, Sidoli:05}, or, in some cases, the central mass of a He white dwarf. \citet{IntZand:05} have shown that the latter donor type is consistent with both the properties of the long X-ray burst observed from the ultracompact binary 2S 0918-549 and that systems enhanced (and constant) Ne/O ratio. 

A third possibility we now propose is that the donors in these ultracompact systems are He white dwarfs, as in 4U 1820-30, and that the accretion of helium rich matter results in RE bursts and ejection of ashes with highly non-solar abundances of Ne and O. Radius expansion bursts have indeed been seen in two bursts from 2S 0918-549 \citep{Cornelisse:02, IntZand:05}. Though only a single realization, the Ne/O ratio by number at $y = y_{\rm wind}$ for model He0.1 shown in Figure \ref{fig:abundyc1} is $\sim 1.0$, comparable to that seen in the four systems. In order for this explanation to work, the ejected ashes must either somehow remain in the environment of the binary or be continuously replenished by periodic RE bursts. The variations in the observed Ne/O ratio in observations of 4U 1543-624 and 4U 1850-087 could just reflect the time since the last RE burst.

Finally, we note that ash ejection may also occur during superbursts that undergo photospheric radius expansion. The amount of mass ejected by superbursts may be much larger than ordinary RE bursts, suggesting even higher column densities of ejected ashes. Furthermore, the ejected ashes are likely to be heavier since the carbon that fuels the superburst lies underneath the heavy rp-ashes of normal bursts.

\acknowledgements 

We thank A. Heger for sharing calculations that encouraged the start
of this study, F.-K. Thielemann for providing the network solver,
A. Sakharuk for fitting and implementing the nuclear reaction rates,
and E. Kuulkers for bringing to our attention earlier
observations. This research was supported by NASA grant NNG05GF69G,
and NSF grants PHY99-07949, AST02-05956, and PHY0216783 (Joint
Institute for Nuclear Astrophysics). H. S. is also supported by NSF
grant PHY 0110253.

\end{document}